\documentclass[12pt]{l4dc2023}
\usepackage{graphics} 
\usepackage{mathptmx} 
\usepackage{times} 
\usepackage{amsmath} 
\usepackage{amssymb, amsfonts}  
\usepackage{newtxtext, newtxmath}
\usepackage{xcolor}
\usepackage{tikz}
\usepackage{enumerate}
\usepackage{newtxmath}
\usepackage{diagbox}
\usepackage{multicol}
\usepackage{multirow}
\usepackage{wrapfig}
\usepackage{url}

\title[Nonlinear State Observer based on Neural ODEs]{Learning Robust State Observers using Neural ODEs (longer version)}
\usepackage{times}

\author{%
  \Name{Keyan Miao}  \thanks{K.M. is supported by the EPSRC under the grant EP/T517811/1. 
 }
 \Email{keyan.miao@eng.ox.ac.uk}\\
 \Name{Konstantinos Gatsis} \Email{konstantinos.gatsis@eng.ox.ac.uk}\\
 \addr Department of Engineering Science, University of Oxford%
}
\begin{document}

\maketitle

\begin{abstract}
Relying on recent research results on Neural ODEs, this paper presents a methodology for the design of state observers for nonlinear systems based on Neural ODEs, learning Luenberger-like observers and their nonlinear extension (Kazantzis-Kravaris-Luenberger (KKL) observers) for systems with partially-known nonlinear dynamics and fully unknown nonlinear dynamics, respectively. In particular, for tuneable KKL observers, the relationship between the design of the observer and its trade-off between convergence speed and robustness is analysed and used as a basis for improving the robustness of the learning-based observer in training. We illustrate the advantages of this approach in numerical simulations.

\end{abstract}

\section{Introduction}
The emergence of deep artificial neural network  architectures (DNNs) as powerful function approximations in machine learning and computer vision tasks has generated an interest in using DNNs, and data-driven approaches more generally, in complex problems that involve systems with dynamics, such as for the purpose of control \citep{DRL}. Finding suitable methodologies for incorporating and training DNNs for systems with dynamics is an important challenge. In this paper, we explore the use of the recent tool of Neural Ordinary Differential Equations (NODEs) for the design of NNs for dynamical systems. Specifically, we consider the task of designing state estimators/observers for systems with nonlinear dynamics.\\
Observers are used to estimate the unmeasured state of a dynamical system based on its output, and they are a key component of closed loop control systems. For linear systems, a general approach is based on the Luenberger observer design originally presented by \cite{Luenberger}. In contrast, there are not many general approaches to observer design for nonlinear systems, a review of which is given in \cite{reviewobserver}. High-gain observers (HGO) \citep{HGO} and Extended Kalman Filter (EKF) \citep{EKF} are most commonly used. 
However, HGO show poor transient performance and a high sensitivity to noise, while EKF only guarantee local convergence. In addition, Luenberger method has also been extended to nonliner systems, leading to the so-called Kazantzis-Kravaris-Luenberger (KKL) observer \citep{KKL}. The idea of this   approach is first immersing the nonlinear system into a latent linear system of higher dimension with an output injection and then mapping this linear system to a state estimate. The existence and injectivity of this mapping is guaranteed by mild observability conditions, which makes this design relatively general. Exponential convergence of KKL observer can be guaranteed by appropriately designing the observer dynamics to have a contraction property and is tuneable under additional observability conditions \citep{convergence}. However, the main challenge lies in computing the latent linear system and the mapping between it and the nonlinear system.\\
To overcome this challenge, an initial attempt at neural network-based KKL observer design is reported by \cite{numericalKKL}, and follow-up research were proposed within the last year (\cite{discreteKKL}, \cite{gaintuning}, \cite{PDE-KKL}). In general this approach trains a neural network to approximate the mapping and its left inverse, using supervised learning methods by fixing the linear dynamics of the KKL observer or using auto-encoders.\\
In this paper, to design nonlinear systems observers, we propose the tool of Neural ODEs. This tool was popularized by \cite{NODE}, which used it to abstract the hidden layers of DNNs as dynamics of a continuous-time system described by an ODE. This approach sparked significant interest in the machine learning community 
recently (e.g. \cite{latent}, \cite{disNODE}, \cite{HNODEs}, \cite{LyaNet}, \cite{TLNODE}, \cite{KidgerOn}). Similar ideas have been discussed by \cite{WeinanProposal} and \cite{qianxiao}.
But beyond static machine learning problems, 
it is natural to use Neural ODEs to solve problems in dynamical systems since dynamical systems are typically represented by ODEs, and initial studies using Neural ODEs for system identification were recently proposed (\cite{Node-dynamics-PINN}, \cite{node-recg-sysid}).\\
The contributions of our paper are: a) we propose a complete approach to the design of state observers for nonlinear systems using Neural ODEs, b) we show the relationship between the paramaters of tuneable KKL observer and its rate of convergence and robustness to model uncertainty and measurement noise, and c) we incorporate this relationship in the training process as a way to design robust observers and show numerically the advantages of this approach compared to the literature. With respect to the literature on learning nonlinear observers, our approach based on Neural ODEs allows us to tune both the latent observer dynamics and its mapping to enhance robustness. With respect to the Neural ODEs literature mentioned above, our paper expands the use of this tool from static machine learning problems to problems when DNNs interact in a closed loop with physical system dynamics.
After providing preliminaries on state observers for nonlinear systems and Neural ODEs in Section \ref{2.1} and \ref{2.2}, we state the problems addressed in this paper and the methodology in Section \ref{3}. Numerical results are presented in Section \ref{4}, showing the benefits of our approach and the limitations particularly in terms of generalization, and Section \ref{5} concludes with conclusions and future perspectives.\\
\textit{Notation}: Throughout this paper, for a vector or a matrix, $\lvert \cdot \rvert$ denotes (induced) norm 2; $\Re$ denotes the real part.
\section{Problem formulation for Nonlinear State Observer}
\label{2.1}
Consider an autonomous nonlinear system:
\begin{equation}
\dot x\left(t\right) = f\left(x\left(t\right)\right),\ y\left(t\right) = h\left(x\left(t\right)\right)
\label{auto}
\end{equation}
with state $x \in \mathbb{R}^{n_x}$, output $y \in \mathbb{R}^{n_y}$, and maps $f: \mathbb{R}^{n_x} \rightarrow \mathbb{R}^{n_x}$ and $h: \mathbb{R}^{n_x} \rightarrow \mathbb{R}^{n_y}$ sufficiently smooth. Our aim is to design a state observer that estimates the value of the state $x$ from the output $y$, and makes the estimated value $\hat x$ converge to the real state value $x$. The  observer of this system takes the general form:
\begin{equation}
\dot z\left(t\right) = \mathcal{F}\left(z\left(t\right), y\left(t\right)\right),\ \hat x\left(t\right) = \mathcal{G}\left(z\left(t\right)\right)
\label{general-observer}
\end{equation}
with state $z \in \mathbb{R}^{n_z}$ (the internal latent state of the observer), output $y \in \mathbb{R}^{n
_y}$, and maps $\mathcal{F}: \mathbb{R}^{n_z}\times \mathbb{R}^{n_y} \rightarrow \mathbb{R}^{n_z}$ and $\mathcal{G}: \mathbb{R}^{n_z} \rightarrow \mathbb{R}^{n_x}$ which need to be designed.
Unlike linear system where Luenberger observers \citep{Luenberger} are wildly recognized, there are not many general approaches to observer design for nonlinear systems.  In a seminal paper, \cite{KKL} have proposed to extend the Luenberger observer for linear systems to the nonlinear case -- called the KKL observer. \cite{KKL2} proved the existence of such an observer under the following conditions. 
\newtheorem{assumption}{Assumption}[section]
\begin{assumption}
\label{asm1}
There exists a compact set $\mathcal{X}$ such that for any solutions $x$ to \eqref{auto} of interest, $x\left(t\right)\in \mathcal{X}$ for all $t \geq 0$.
\end{assumption}
\begin{assumption}
\label{asm2}
There exists an open bounded set $\mathcal{O}$ containing $\mathcal{X}$ such that \eqref{auto} is backward $\mathcal{O}$-distinguishable on $\mathcal{X}$, namely for any trajectories $x_a$ and $x_b$ of \eqref{auto} such that  $\left(x_a\left(0\right), x_b\left(0\right)\right) \in \mathcal{X} \times \mathcal{X}$ and $x_a\left(0\right) \neq x_b\left(0\right)$, there exists $t \leq 0$ such that 
$
h\left(x_a\left(t\right)\right) \neq h\left(x_b\left(t\right)\right)
$
and $\left(x_a\left(\tau\right), x_b\left(\tau \right)\right) \in \mathcal{O} \times \mathcal{O}$ for all $\tau \in \left[t, 0\right]$. In other words, their respective outputs become different in backward finite time 
before leaving $\mathcal{O}$.
\end{assumption}
\begin{theorem}[\cite{KKL2}]
\label{thm1}
Suppose Assumptions \ref{asm1} and \ref{asm2} hold. Define $d_z = d_y\left(d_x + 1\right)$, then there exists $l > 0$ and a set $\mathcal{S}$ of zero measure in $\mathbb{C}^{d_z}$ such that for any matrix $D \in \mathbb{R}^{d_z\times d_z}$ with eigenvalues $\left(\lambda_1, ..., \lambda_{d_z}\right)$ in $\mathbb{C}^{d_z} \setminus \mathcal{S}$ with $\Re\lambda_i < - l$, and any $F \in \mathbb{R}^{d_z \times d_x}$ such that $\left(D,  F\right)$ is controllable, there exists an injective mapping $\mathcal{T}: \mathbb{R}^{d_x} \rightarrow \mathbb{R}^{d_z}$ and a pseudo-inverse $\mathcal{T}^*: \mathbb{R}^{d_z} \rightarrow \mathbb{R}^{d_x}$ such that the trajectories of \eqref{auto} remaining in $\mathcal{X}$ and any trajectory of
\begin{equation}
    \dot z\left(t\right) = Dz\left(t\right) + Fy\left(t\right)
\label{KKL-z}
\end{equation}
satisfy
\begin{equation}
\lvert z\left(t\right) - \mathcal{T}\left(x\left(t\right)\right)\rvert \leq M \lvert z\left(0\right) - \mathcal{T}\left(x\left(0\right)\right)\rvert e^{\lambda_{\max}t},\ \lambda_{\max} = \max \{\Re \lambda_1, ..., \Re \lambda_{d_z}\}
\end{equation}
for some $M > 0$, and $\mathcal{T}$ is the solution of the partial differential equation
\begin{equation}
\frac{\partial \mathcal{T}}{\partial x}\left(x\right)f\left(x\right) = D\mathcal{T}\left(x\right) + Fh\left(x\right)
\label{PDE}
\end{equation}
and
\begin{equation}
    \lim_{t\to +\infty}\lvert x\left(t\right)-\mathcal{T}^*\left(z\left(t\right)\right)\rvert = 0
\end{equation}
\end{theorem}
The linear dynamics \eqref{KKL-z} and the mapping $\hat{x} = \mathcal{T}^*(z)$ define the KKL observer, which is a special case of \ref{general-observer}. Although the above theorem guarantees the observer exists in a wide class of systems, it is still very difficult to find such $\mathcal{T^*}$. Additionally, there are many tuning choices that affect the performance of the observer, for example the matrix $D$ affects convergence and sensitivity to measurement noise \citep{gaintuning}, although the relationship between performance and variables was not made explicit.
\section{Preliminaries on Neural ODEs}
\label{2.2}
It has been shown that DNNs can learn nonlinear mappings and can generalize to unseen data under certain conditions for a diverse range of problems. 
At one time, neural networks faced bottlenecks due to the limitations of the depth problem. The impetus for this breakthrough was the Residual Network (ResNet) proposed by \cite{ResNet} which added skip connections to the network, resulting in a remarkable improvement in the performance of the network. Its universal approximation power has recently been explained from a non-linear control perspective \citep{uni-appro}. Meanwhile, the idea of treating the hidden layers of neural networks as states of a dynamical system became popular when ResNet was proposed.
\cite{NODE} proposed an ODE specified by the neural network to parameterize the continuous dynamics:
\begin{definition}[Neural ODEs]
With $h_x: \mathbb{R}^{n_x} \rightarrow \mathbb{R}^{n_z}$, $h_y: \mathbb{R}^{n
_z} \rightarrow \mathbb{R}^{n_y}$ representing the input network and output network respectively, a Neural ODE is a system of the form
\begin{equation}
\left\{
\begin{array}{ll}
\dot {z} \left(t\right) = f\left(t, z\left(t\right), \theta\right) & \\
z\left(t_0\right) = h_x\left(x\right) & t \in \mathcal{S} \\
\hat {{y}} \left(t\right)=h_y\left(z\left(t\right)\right)
\end{array}
\right.
\label{node}
\end{equation}
where $\mathcal{S}:= \left[t_0, t_f\right]$ ($t_0, t_f\in \mathbb{R}^+$) is the depth domain and $f$ is a neural network called ODENet which is chosen as a part of the machine learning model with parameter $\theta$.
\end{definition}
The solution of this ODE at some time $t_f$ from an initial value $z\left(t_0\right)$ is $z\left(t_f\right)$, obtained with a differential equation solver by means of a specific solution scheme according to desired accuracy. The number of times the ODE solver evaluates the function in one forward pass can be interpreted as the number of hidden layers of the neural network, i.e.,, the depth. Now ResNet can be seen a special case of Neural ODEs  using Euler discretization.\\
For common machine learning scenarios, such as image classification, an input data point $x$ is mapped by Neural ODEs to $\hat{y}(t_f)$, e.g., a label. In other words, the inference of Neural ODEs is carried out by solving the initial value problem (IVP):
\begin{equation}
\hat y\left(t_f\right) = h_y\left(h_x\left(x\right)+\int_{t_0}^{t_f} f\left(t, z\left(t\right),\theta\right)dt\right)
\end{equation}
The training of the Neural ODEs with parameters $\theta$ proposed in \cite{NODE} considers only a  loss function that depends on the terminal state $\hat y\left(t_f\right)$, which is also the common scenario for supervised learning problems where the terminal state is compared to a true label. In this case, the training can be cast into a \textit{Mayer} optimal control problem. However, for the purpose of our paper, it is valuable to introduce 
\begin{equation}
\ell \coloneqq \Phi\left(x \left(t_f\right)\right) + \int_{t_0}^{t_f}L\left(z\left(t\right),\theta,t\right)dt
\label{int}
\end{equation}
since in the framework of Neural ODEs, the latent states evolve through layers, and then generate outputs.
With such a loss function, the training can be cast into an optimization problem of the form
\begin{equation}
\begin{aligned}
& \quad \quad \min_{\theta \in U} \quad \ell\\
s.t. & \quad \dot {z} \left(t\right) = f\left(t, z\left(t\right), \theta\right), t \in \mathcal{S}\\
& \quad z\left(t_0\right) = h_x\left(x\right), \hat {y} \left(t\right)=h_y\left(z\left(t\right)\right)
\end{aligned}
\label{bolza}
\end{equation}
which is a \textit{Bolza} optimal control problem \eqref{bolza} \citep{MP}, and the problem can be solved recursively by gradient descent (GD). 

\section{Learning  Nonlinear State Observers using Neural ODEs}
\label{3}
We now return to our main problem of learning state observers of the form \eqref{general-observer} for systems described in \eqref{auto}. We propose the general framework of solving the state observation problem of nonlinear systems based on Neural ODEs:
\begin{equation} 
\begin{aligned}
\dot z\left(t\right) &= \mathcal{F}\left(z\left(t\right), y\left(t\right), \theta\right),\ \hat x\left(t\right) =\mathcal{G}\left(z\left(t\right), \eta\right)\\
\mbox{s.t.\quad \quad} \dot x\left(t\right) &= f\left(x\left(t\right)\right),\ y\left(t\right) = h\left(x\left(t\right)\right)
\end{aligned}
\label{unknown}
\end{equation}
where $\hat x$ is the estimated state. The ODE that describes the dynamics of latent state $z \left(t\right)$ is represented by a neural network $\mathcal{F}$ whose parameters are $\theta$, taking $z \left(t\right)$ and $y\left(t\right)$ as input. Transformation $\mathcal{G}$ parameterized by $\eta$ can be seen as the output network stated in Neural ODEs structure \eqref{node}.
Furthermore, we propose the following loss function   to be minimized for \eqref{unknown}
\begin{equation}
    \ell \coloneqq \int_{t_0}^{t_f} L dt = \int_{t_0}^{t_f} \lvert x\left(t\right) - \hat x\left(t\right) \rvert^2 dt
\label{loss}
\end{equation}
which is a \textit{Lagrange} optimal control problem. Gradients of $\ell$ with respect to parameter $\theta$ can be computed via automatic differentiation or adjoint sensitivity analysis as follows which is with only $\mathcal{O}(N_f)$ memory cost during training \citep{ACA} where $N_f$ denotes the number of layers of $\mathcal{F}$.  $\mathcal{F}$ can incorporate a priori knowledge, such as physical information about the system, and the theory of nonlinear systems observer design, as described below.
\begin{proposition}
\label{pro1}
Consider the problem \eqref{unknown}-\eqref{loss}, the gradient of loss $\ell$ with respect to parameter $\theta$ is
\begin{equation}
\nabla_\theta \ell = \mu \left(t_0\right)
\label{grad}
\end{equation}
where $z\left(t\right)$, $p\left(t\right)$ and $\mu\left(t\right)$ satisfy the boundary value problem:
\begin{equation}
\begin{aligned}
& \dot {z}\left(t\right) = \mathcal{F}, \; z\left(t_0\right) = z_0\\
& \dot p\left(t\right) = -p\left(t\right) \frac{\partial \mathcal{F}}{\partial z} - \frac{\partial L}{\partial z}, \; p\left(t_f\right) = \vmathbb{0}_{n_{z}}\\
& \dot {\mu} \left(t\right) = -p\left(t\right)\frac{\partial \mathcal{F}}{\partial \theta} - \frac{\partial L}{\partial \theta}, \; \mu \left(t_f\right) = \vmathbb{0}_{n_{\theta}}
\end{aligned}
\end{equation}
\end{proposition}
Detailed derivations are provided in Appendix \ref{profor3}. With the gradients with respect to the ODE parameters $\theta$ computed,
the training process based on GD then can be carried out. The training data is generated by solving the dynamics with a numerical ODE solver over a finite time interval from many initial conditions, chosen randomly from a Gaussian distribution, and during the generation, output $y(t)$ is stored for the training as the input of ODENet. In our numerical experiments, it is possible to sample the data and train the model simultaneously for convenience, in which case the states of Neural ODEs can be extended as a stack of $z$ with 
$x$, and the dynamics of $x$ comes from the system rather than the network. This means that \textit{our training does not require knowledge of the nonlinear system dynamics}, however it requires measurements of the true system states and outputs.
\tikzset{every picture/.style={line width=0.75pt}} 
\begin{figure}[ht]
 \vspace{-10pt}
\centering
\input{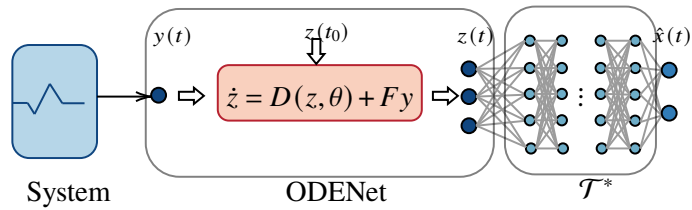}
\vspace{-10pt}
\caption{Schematics of the learned KKL Observer with Neural ODEs}
\label{KKLscheme}
\vspace{-20pt}
\end{figure}
\subsection{Luenberger-like Observer for Systems with Partially-known Dynamics based on Neural ODEs}
For systems whose linear output function $C$ and the linear part $A$ of dynamics are known, the Luenberger-like observer can be designed as \eqref{node-partial} show. We incorporate the prior knowledge of system into Neural ODEs framework and take advantage of the strong ability of neural networks to represent nonlinear functions, here we use a three-layer network $\hat g$ parameterized by $\theta$ to learn the unknown nonlinear part \citep{obs-lyapunov}. $G$ is the observer gain selected such that the matrix $A-GC$ is a Hurwitz matrix. The framework of Neural ODEs observer for this problem is proposed as:
\begin{equation} 
\begin{aligned}
\dot {\hat x}\left(t\right) &= A \hat x\left(t\right) + \hat g\left(\hat x\left(t\right), \theta\right) + G\left(y\left(t\right) - \hat y\left(t\right)\right),\ \hat y\left(t\right) = C \hat x\left(t\right)\\
\mbox{s.t.\quad \quad} \dot x\left(t\right) &= Ax\left(t\right) + g\left(x\left(t\right)\right), \quad y\left(t\right) = C x\left(t\right)
\end{aligned}
\label{node-partial}
\end{equation}
Based on Neural ODEs stated in \eqref{node}, in this framework, the input network $h_x$ and output network $h_y$ can both be interpreted as identity transformations. 
\subsection{KKL observer for Systems with Unknown Dynamics based on Neural ODEs}
When both the nonlinear system dynamics and the output measurement mappings are completely unknown given by \eqref{auto}, we propose to use the Neural ODEs framework to learn KKL observers as defined in Section \ref{2.1}, learning the matrix $D$ and the mapping $\mathcal{T^*}$ jointly. To ensure that the matrix $D$ is Hurwitz, it is designed as a diagonal matrix and $\theta$ represents the values on the diagonal, i.e., the eigenvalues of $D$, restricted to be negative real numbers. $F$ is designed as $F = \boldsymbol{1}_{d_z\times d_y}$ to guarantee the controllability of $\left(D, F\right)$. Let $\hat{\mathcal{T}}^*$ represent the neural network parameterized by $\eta$ that approximates $\mathcal{T}^*$. The problem is stated as \eqref{KKLobserver} where $z\left(t\right)$ is the latent state whose dynamics is described by ODENet.
\begin{equation} 
\begin{aligned}
\dot z\left(t\right) &= D\left(z\left(t\right), \theta\right) + Fy\left(t\right) = Diag\{\theta_1,...,\theta_{d_z}\}z\left(t\right) + Fy\left(t\right)\\
\hat x\left(t\right) &= \hat{\mathcal{T}}^*\left(z\left(t\right), \eta\right)\\
\mbox{s.t.\quad \quad} \dot x\left(t\right) &= f\left(x\left(t\right)\right),\ y\left(t\right) = h\left(x\left(t\right)\right)
\end{aligned}
\label{KKLobserver}
\end{equation}
The schematics of the Neural ODEs observer incorporating KKL is illustrated in Figure \ref{KKLscheme}.
Based on Neural ODEs stated in \eqref{node}, in this framework, the input network $h_x$ can be interpreted as an identity transformation and the output network is $\hat{\mathcal{T}}^*$. We propose the loss function defined as \eqref{loss}
which penalizes the error between estimated state and true state value. Finally, our approach can be extended to non-autonomous systems by applying a stationary transformation \citep{kkl-non} that does not require a specific, well-chosen excitation that \cite{numericalKKL} proposed.
\subsubsection{Robustness of KKL observer}
The aim of this subsection is to characterize robustness, i.e., what happens if we set up the design of the state observer as in section \ref{2.1}, but then when deploying the state observer onto the system, the latter is subject to model uncertainties and measurement moise  that will impact the state observer. 
Consider  the case where noise in the system dynamics and measurements is present, we model a nonlinear system
\begin{equation}
\dot x\left(t\right) = f\left(x\right) + w\left(t\right), y\left(t\right) = h\left(x\left(t\right)\right) + v\left(t\right)
\label{noise}
\end{equation}
where $w \in \mathbb{R}^{n_x}$ and $v \in \mathbb{R}^{n_y}$. They represent model uncertainties and measurement noise respectively which are unknown but bounded signals such that $\sup_{t\in\mathcal{S}}\lvert w(t)\rvert \leq \bar w$, $\sup_{t\in\mathcal{S}}\lvert Fv(t)\rvert \leq \bar v$. Here, we focus on how the choice of $D$ will affect the robustness of the observer by scaling its eigenvalues.
\begin{proposition}
\label{robust}
Suppose Assumptions \ref{asm1} and \ref{asm2} hold and assume infinitesimal distinguishability property holds \citep{convergence}. Consider the mapping $\mathcal{T}$ obtained by \eqref{PDE} where matrix $D = Diag\{\lambda_1,...,\lambda_{d_z}\}$. Assume there exist positive real numbers $L_\mathcal{T}>0$ and $L_{\mathcal{T}_x}>0$ such that for all $x \in \mathcal{X}$
\begin{equation}
\lvert x_1 - x_2 \rvert \leq L_\mathcal{T} \lvert \mathcal{T}\left(x_1\right) - \mathcal{T}\left(x_2\right) \rvert,\ 
\label{assum} \qquad
\text{and}\quad \lvert \frac{\partial \mathcal{T}}{\partial x}\left(x\right)\rvert \leq L_{\mathcal{T}_x}.
\end{equation}
For any Hurwitz matrix $D_k = kD$, which means that all the eigenvalues are multiplied by a positive real number $k \geq 1$, let the corresponding mappings be $\mathcal{T}_k$ and $\mathcal{T}_k^*$ obtained by solving $\frac{\partial \mathcal{T}_k}{\partial x}\left(x\right)f\left(x\right) = D_k\mathcal{T}_
k\left(x\right) + Fh\left(x\right)$. Then the trajectory $\{x\left(t\right), \tilde x\left(t\right)\}$, where $x$ represents the state of system \eqref{noise} and 
$\tilde x = \mathcal{T}_k^* \left(z\right)$ where the dynamics of $z$ is $\dot z\left(t\right) = D_k z\left(t\right) + Fy\left(t\right)$, satisfies
\begin{equation}
\lvert x\left(t\right) - \tilde x\left(t\right) \rvert \leq k^{n_z} \sqrt n_x L_\mathcal{T} \exp\left(k \lambda_{max}t\right)\lvert z\left(t_0\right) - \mathcal{T}_k\left(x\left(t_0\right)\right)\rvert + \frac{k^{n_z}\sqrt n_x L_\mathcal{T} }{\lvert k \lambda_{max}\rvert}\left(\frac{L_{\mathcal{T}_x}}{k}\bar w + \bar v\right)
\end{equation}
\end{proposition}
This proposition demonstrates the trade-off between convergence speed and robustness to model uncertainties and measurement noise: when increasing the coefficient $k$, i.e., enlarging the eigenvalues, the observer's convergence speed increases, but its robustness to model uncertainties and measurement noise decreases. A detailed proof is provided in Appendix \ref{profor4}. 

\begin{remark}
Learning-based KKL Observer Robustness to Approximation Error:
The mapping that is learned by our Neural ODE approach, denoted here as $\hat {\mathcal{T}}^*\left(z\right)$, is only an approximation of $\mathcal{T}^*$. Therefore, the performance of the observer is further affected by the approximation error. Given a Lipschitz continuous activation function such as ReLU, it can be derived that the error between estimated state $\hat x = \hat {\mathcal{T}}^*(z)$ and $x$ is bounded. A detailed proof is provided in Appendix \ref{proforrem}. 
Moreover, the upper bound can be reduced by improving the design and learning techniques of the neural network and by increasing the size of training dataset.
\end{remark}
\subsubsection{Training method to improve robustness}
According to Proposition \ref{robust}, the choice of matrix $D$, or more precisely its eigenvalues, will have a significant impact on the speed of convergence and robustness of the observer, and if no adjustments are made to the training, the learned observer will be concerned only with the speed of convergence. Here, we propose two methods to improve its performance on noise rejection.
\begin{itemize}
    \item Adding noise to the data during training
    \item Using a regularization technique for the parameters (eigenvalues) of $D$ through a penalty added to loss function 
    \begin{equation}
        \tilde \ell \coloneqq \ell + \gamma \ell_{reg} = \ell + \gamma \int_{t_0}^{t_f} \lvert \theta\left(t\right) \rvert^2 dt
    \end{equation}
\end{itemize}
Regularization techniques are often used in machine learning to solve overfitting problems, and poor robustness often implies overfitting. However, it should be clarified that the use of regularization in general machine learning problems is based on the premise that the overfitting problem originates from an overly complex network model, whereas here we use it based on the aforementioned analysis of robustness, i.e., that fast poles can lead to poor robustness of the observer. The addition of noise to the training process is considered to play an equivalent role to regularization \citep{noise+reg}.
\begin{remark}[Comparison with other approaches]
First, in the approach proposed by \cite{numericalKKL}, the matrix $D$ needs to be pre-specified to generate the dataset $\left\{x\left(t_i\right), z\left(t_i\right)\right\}$. Hence, the opportunity to fully tune the observer is missed in prior work, and similarly in \cite{PDE-KKL}. In contrast, our Neural ODEs approach makes it easy to design $D$ and we show numerically the advantages of this approach in the following section.
\cite{gaintuning} proposed a method to improve the choice of $D$ inspired by $\mathcal{H}_{\infty}$ control design, but in practice their method has shortcomings. It is an a posteriori method that requires the mappings corresponding with different $D$ in order to select a better observer whereas their auto-encoder method requires full knowledge of the nonlinear system dynamics, and it does not directly improve robustness. In contrast, our Neural ODE approach is flexible enough to overcome these shortcomings.
\end{remark}
\section{Numerical Results}
\label{4}
\begin{figure}[ht]
\vspace{-10pt}
  \centering
  \includegraphics[scale=0.45]{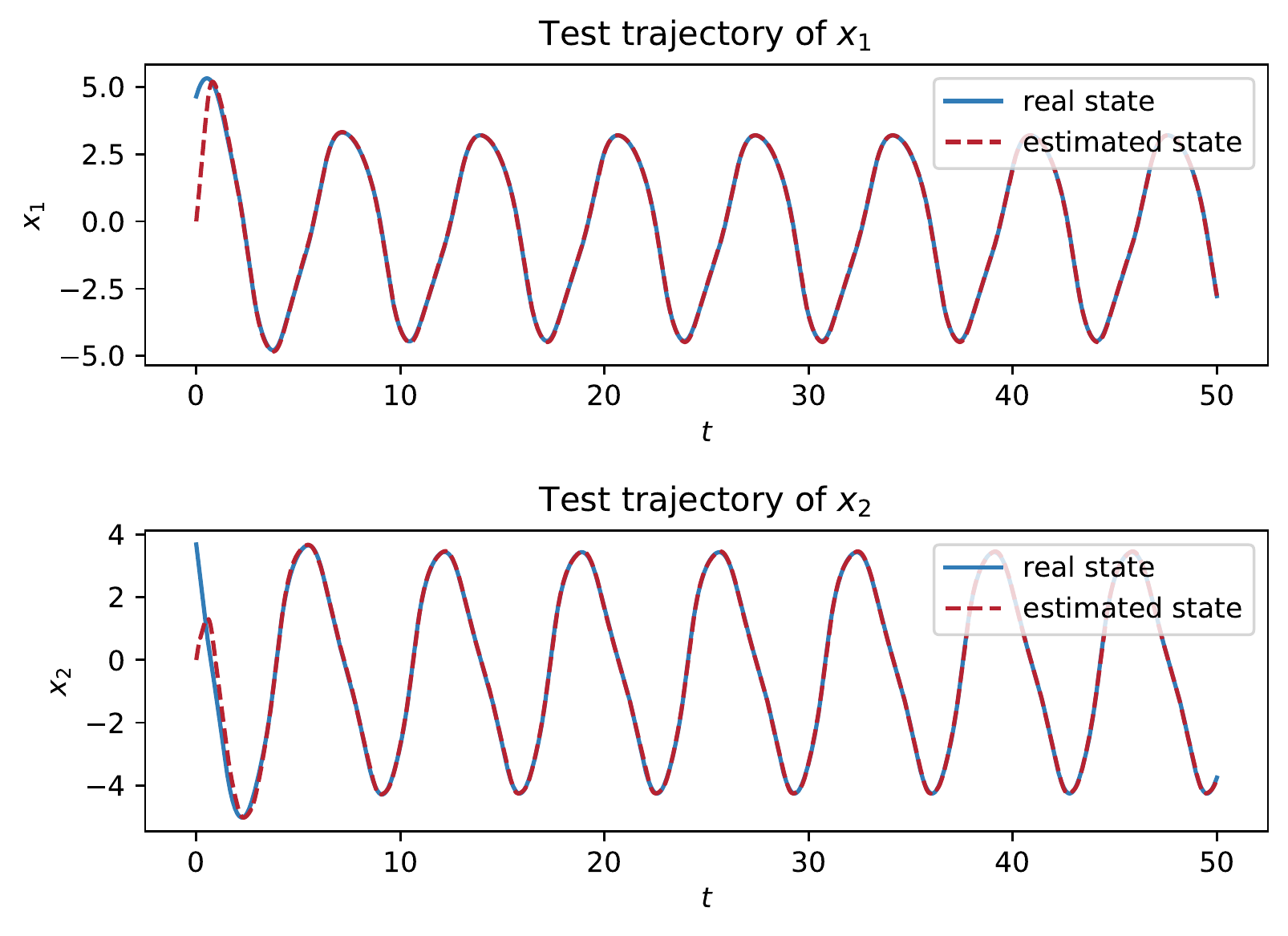}
  \hspace{0.01in}
  \includegraphics[scale=0.45]{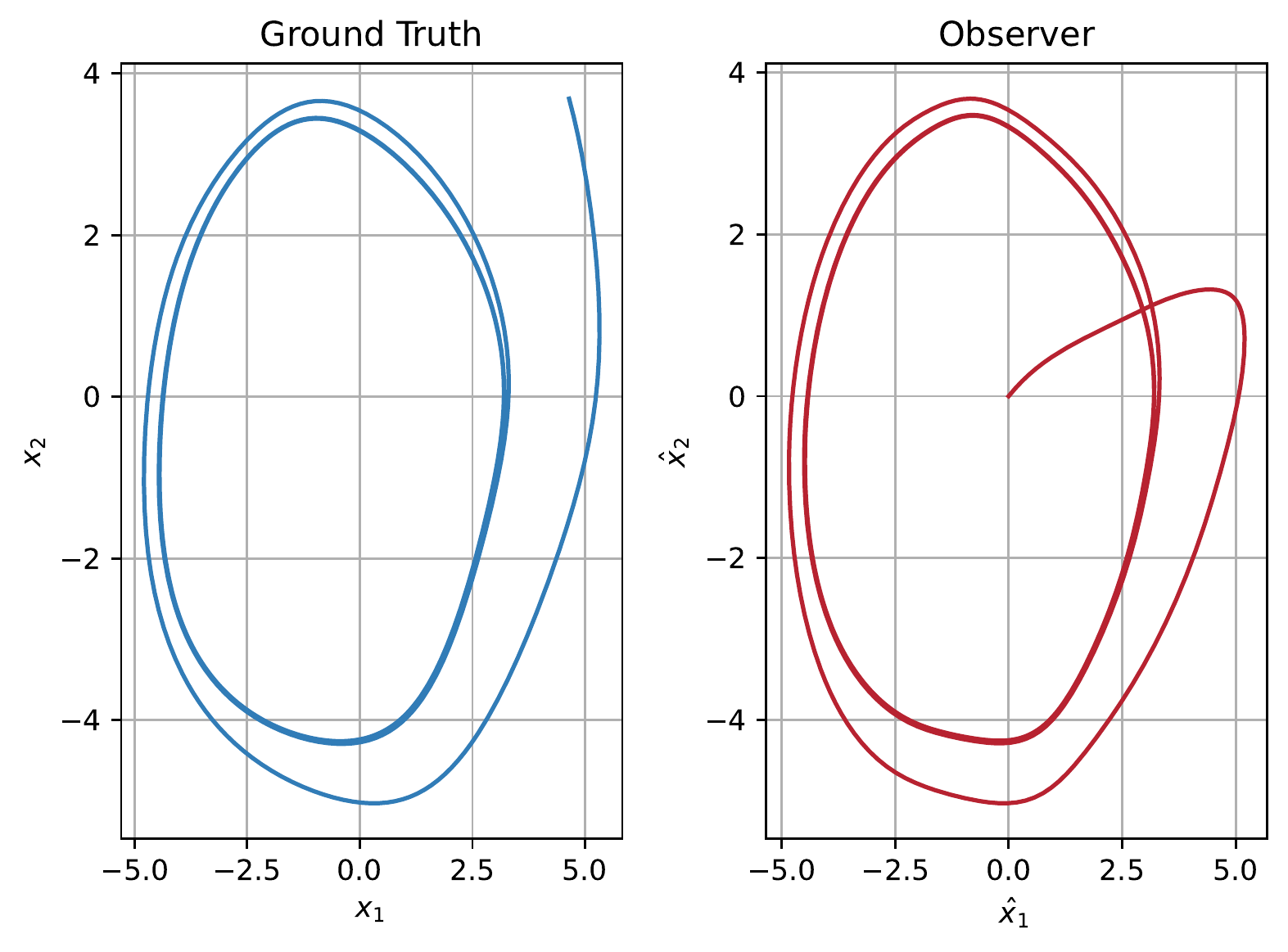}
\vspace{-10pt}
  \caption{NODEs-based Luenberger-like observer result for a given initial condition}
  \label{luen-like}
  \vspace{-10pt}
\end{figure}

\begin{figure}[t]
  \centering
  \includegraphics[scale=0.35]{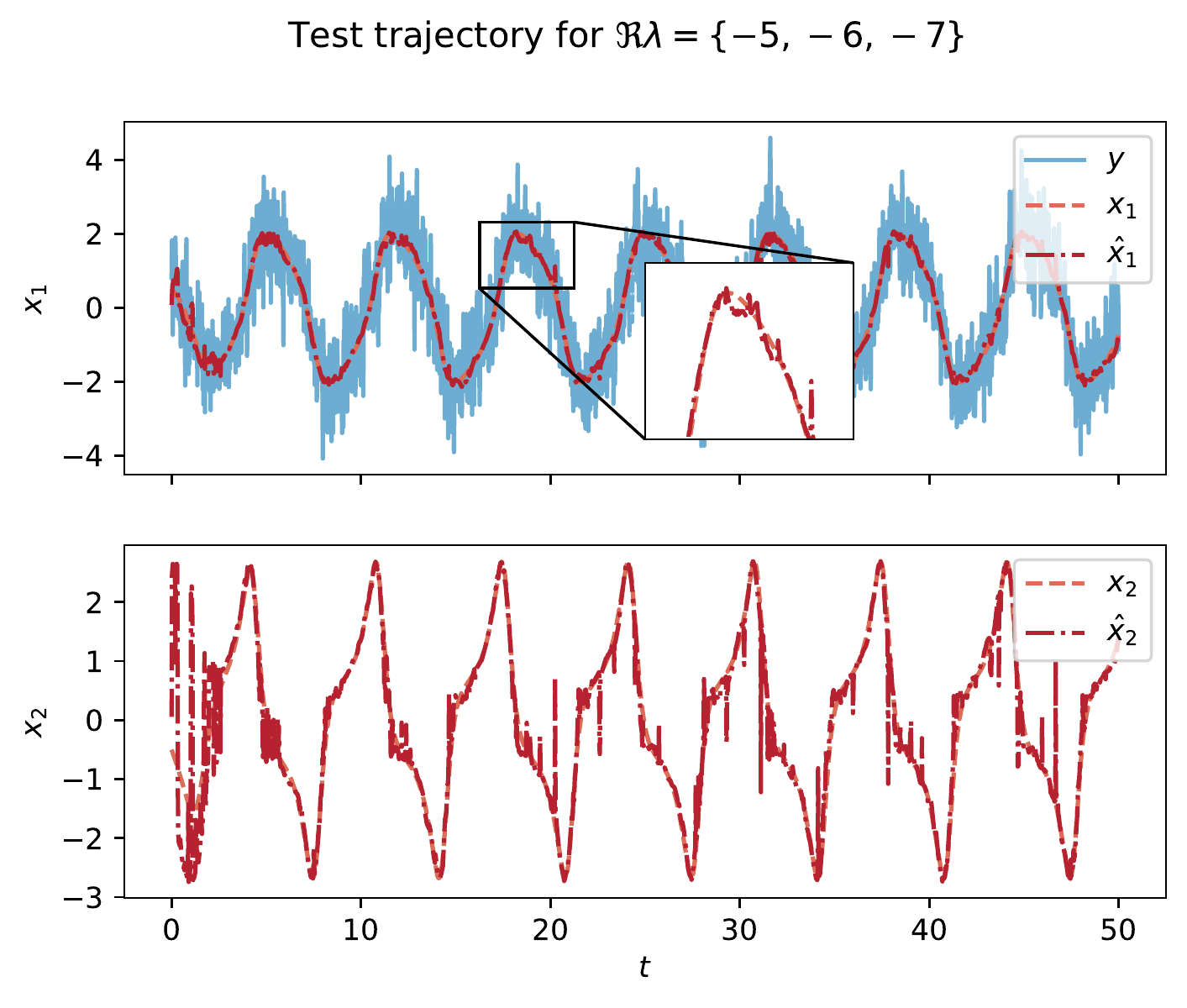}
  \hspace{-0.2cm}
  \includegraphics[scale=0.35]{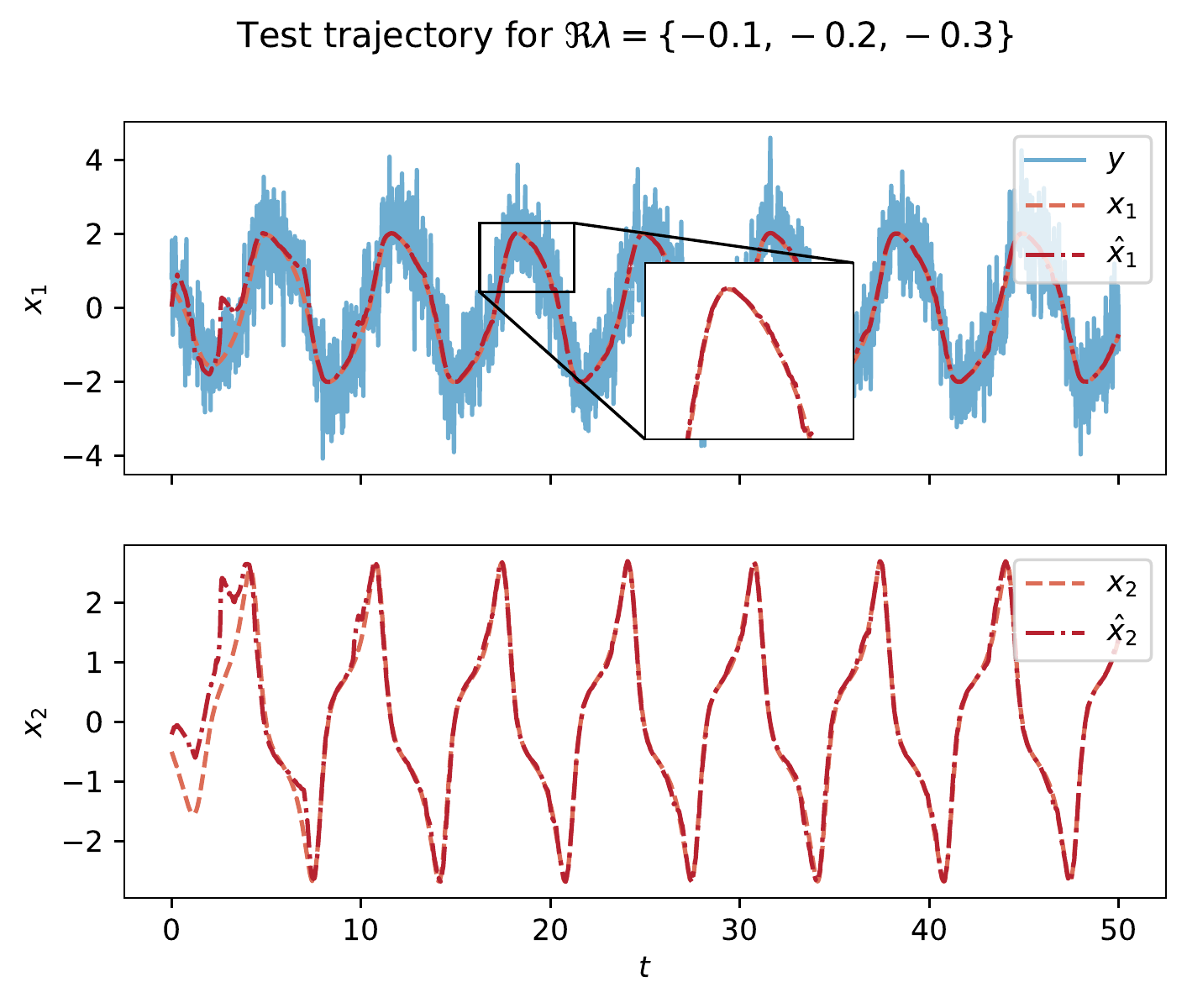}
  \hspace{-0.2cm}
  \includegraphics[scale=0.35]{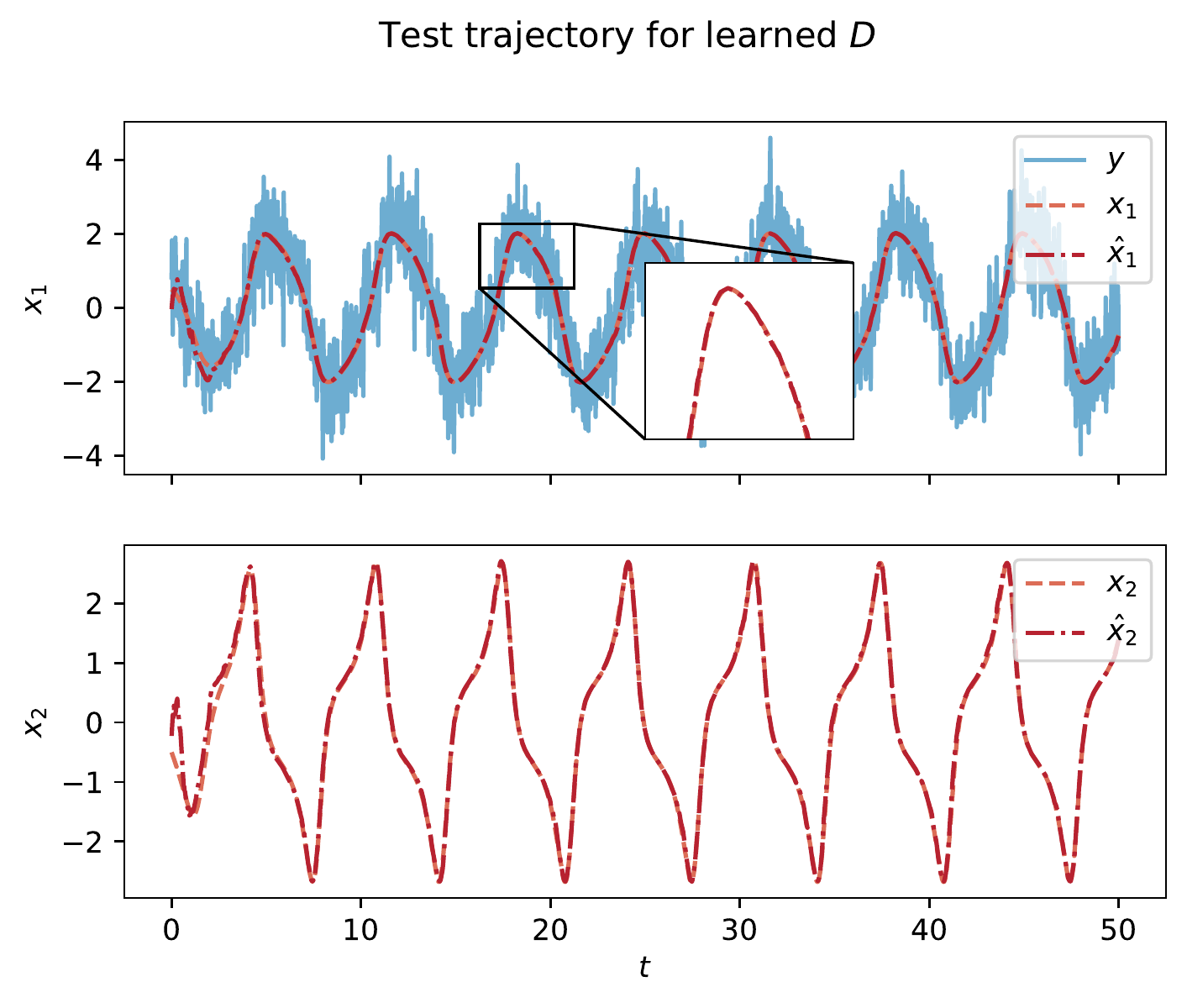}
\vspace{-10pt}
  \caption{Test trajectories of Van der Pol for $x_0 = (-0.5, 0.5)$ with Gaussian noise $\mathcal{N}(0,0.5)$}
  \label{vanderpol-compare}
\vspace{-10pt}
\end{figure}

\begin{figure}[ht]
\vspace{-10pt}
  \centering
  \begin{minipage}[t]{0.5\textwidth}
  \centering
  \includegraphics[scale=0.5]{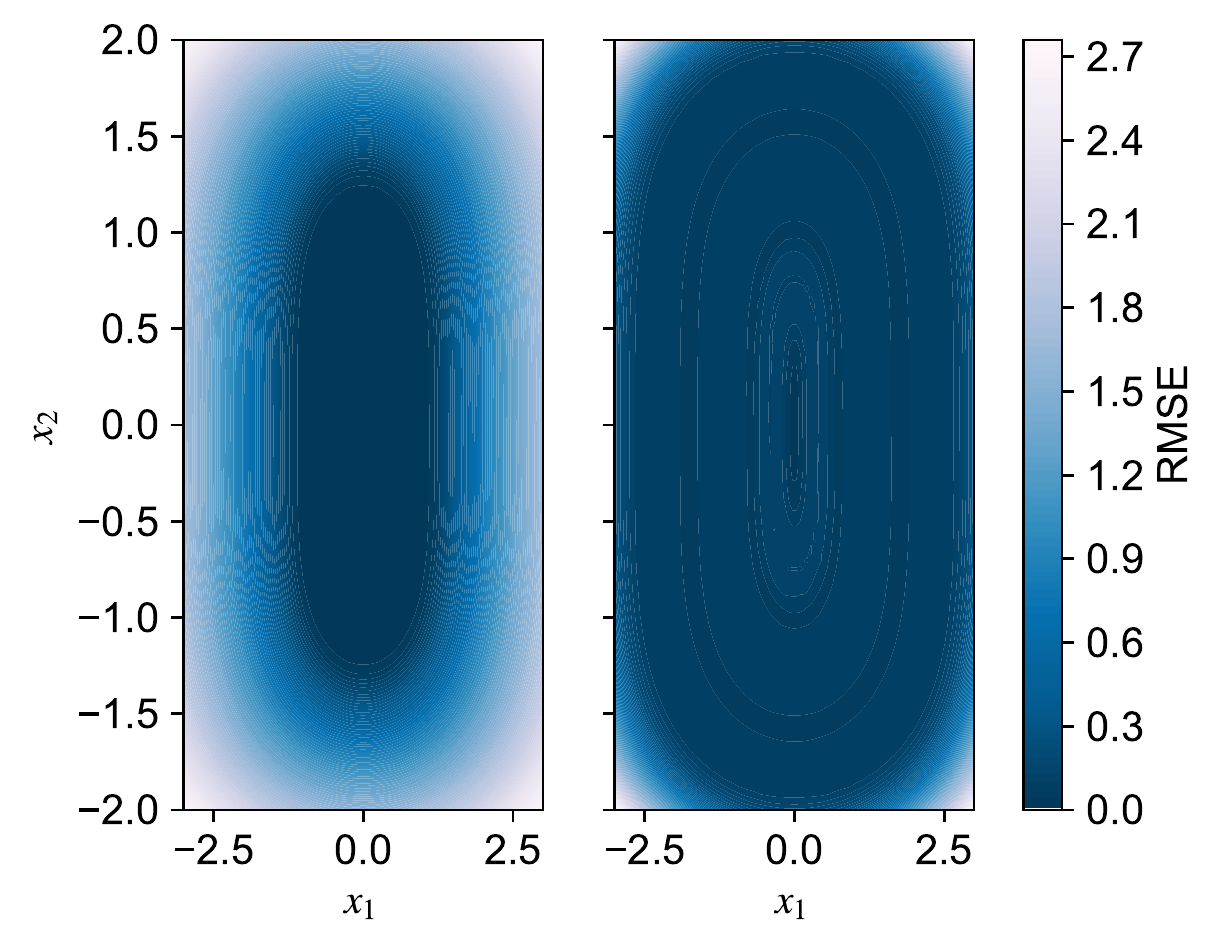}
  \vspace{-20pt}
  \caption{RMSE mapping with different training domain}
  \label{compare_x_rmse}
  \end{minipage}
  \begin{minipage}[t]{0.48\textwidth}
  \centering
  \includegraphics[scale=0.5]{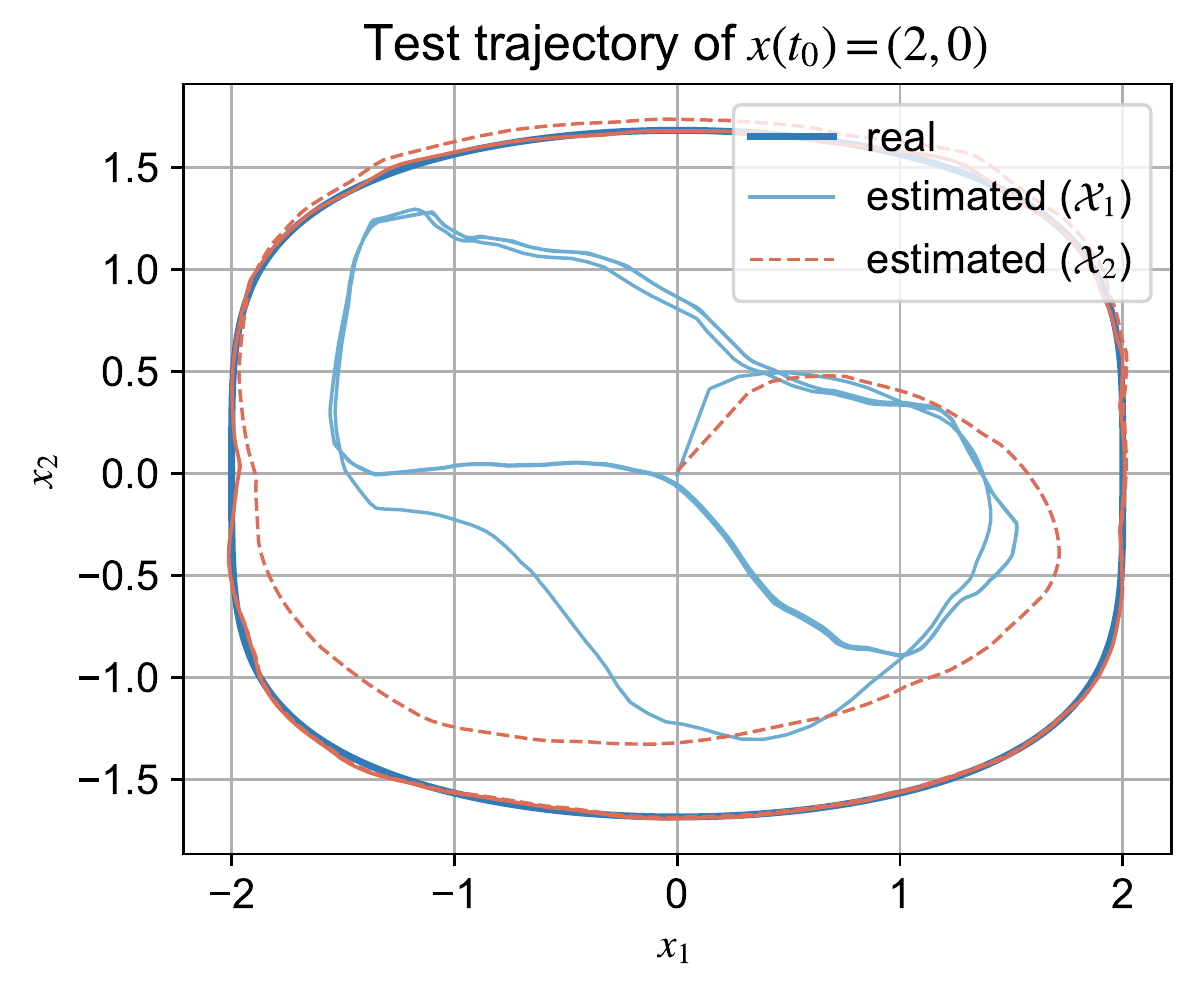}
  \vspace{-20pt}
  \caption{Test trajectory with different training domain}
  \label{compare_x12}
  \end{minipage}
 \vspace{-20pt}
\end{figure}

We illustrate and evaluate our approach through numerical simulations.
\paragraph{Example 1}
Consider an autonomous system 
\begin{equation}
\left\{
\begin{aligned}
\dot x_1\left(t\right) &= x_2\left(t\right) + \sin{x_1\left(t\right)}\\
\dot x_2\left(t\right) &= -x_1\left(t\right) + \cos{x_2\left(t\right)}
\end{aligned}
\right.
\quad y\left(t\right) = x_1\left(t\right)
\end{equation}
where output function is known. Applying the methodology presented in Section \ref{3}, the training data (initial state) is generated from a uniform random distribution $\mathcal{X} = \left[-5, 5\right] \times \left[-5,5\right]$, and the dynamics are solved over the time interval $\left[0, t_f\right]$ with $t_f = 50s$. The resulting observer is shown in Figure \ref{luen-like} for an arbitrary given initial condition. It can be found that the observer exhibits good performance and the error stabilizes after about $5s$.
\paragraph{Example 2: Van der Pol} Consider the autonomous Van der Pol oscillator
\begin{equation}
\left\{
\begin{aligned}
\dot x_1\left(t\right) &= x_2\left(t\right)\\
\dot x_2\left(t\right) &= \left(1-x_1^2\left(t\right)\right)x_2\left(t\right)-x_1\left(t\right)
\end{aligned}
\right.
\quad y\left(t\right) = x_1\left(t\right)
\label{vanderpol}
\end{equation}
which admits a unique limit cycle. To demonstrate how the choice of $D$ will affect the convergence speed and robustness of the KKL observer, we conducted the following experiments respectively:  using the method in \cite{numericalKKL}, specifying the eigenvalues of the diagonal matrix $D$ as $\{-5,-6,-7\}$, $\{-0.1, -6, -7\}$ and $\{-0.1, -0.2, -0.3\}$ as baseline; using the aforementioned NODEs-based scheme, learning the KKL observer with noise added to the training data and using regularization technique respectively. The training data (initial state) is generated from a uniform random distribution $\mathcal{X} = \left[-1, 1\right] \times \left[-1,1\right]$, and the dynamics are solved over the time interval $\left[0, t_f\right]$ with $t_f = 50s$. The learned observers based on our approach and baseline are then tested on arbitrary initial conditions and the root mean square error (RMSE) of different learned observers and their performance are illustrated in Table \ref{vanderpol-rmse} and Figure \ref{vanderpol-compare}.
\begin{table}[t]
\vspace{-10pt}
\centering
\caption{RMSE of learned observers for Van der Pol oscillator tested in different scenarios}
\label{vanderpol-rmse}
\resizebox{.95\columnwidth}{!}{
\begin{tabular}{c|ccc|cc}
\hline
\multirow{2}{*}{\diagbox{Test Scenario}{Observer}} &
  \multicolumn{3}{c|}{Learned with fixed $D$ 
  (\cite{numericalKKL})} &
  \multicolumn{2}{c}{Learned via Neural ODEs (our method)} \\ \cline{2-6} 
 &
  \multicolumn{1}{c|}{\{-5, -6, -7\}} &
  \multicolumn{1}{l|}{\{-0.1, -6, -7\}} &
  \{-0.1, -0.2, -0.3\} &
  \multicolumn{1}{c|}{Learned with Gaussian noise  $\mathcal{N}(0,0.5)$} &
  Learned with Regularization \\ \hline
No Noise &
  \multicolumn{1}{c|}{$\mathbf {0.0548}$} &
  \multicolumn{1}{c|}{$0.1786$} &
  $0.2080$ &
  \multicolumn{1}{c|}{$0.0603$} &
  $0.0712$ \\ \hline
Gaussian Noise $\mathcal{N}(0,0.5)$ &
  \multicolumn{1}{c|}{$0.1160$} &
  \multicolumn{1}{c|}{$0.1903$} &
  $0.2273$ &
  \multicolumn{1}{c|}{$\mathbf{0.0667}$} &
  $0.0863$ \\ \hline
Uniform Noise $\mathcal{U}(-3,3)$ &
  \multicolumn{1}{c|}{$0.3205$} &
  \multicolumn{1}{c|}{$0.2586$} &
  $0.2560$ &
  \multicolumn{1}{c|}{$\mathbf{0.1111}$} &
  $0.1462$ \\ \hline
\end{tabular}}
\vspace{-10pt}
\end{table}
The third observer shown in Figure \ref{vanderpol-compare} is learned with training data added by Gaussian noise. It can be found that when $D$ is specified with small absolute eigenvalues, the observer suffers little from noise but takes about 10 seconds to converge, whereas when a matrix $D$ with larger absolute eigenvalues is chosen, it converges quickly but  is severely affected by noise. In contrast, the observer based on Neural ODEs showed good performance in terms of both convergence speed and robustness. This demonstrates the advantages of our approach over previous approaches.
\paragraph{Example 3: Reverse Duffing Oscillator} Consider the reverse Duffing oscillator
\begin{equation}
\left\{
\begin{aligned}
\dot x_1\left(t\right) &= x_2^3\left(t\right)\\
\dot x_2\left(t\right) &= -x_1\left(t\right)
\end{aligned}
\right.
\quad y\left(t\right) = x_1\left(t\right)
\label{oscillator}
\end{equation}
which admits bounded trajectories where $x_1^2 + x_2^4$ is constant. The size of the training state space subset is critical, and Figure \ref{compare_x_rmse} shows RMSE of results for observers trained under different subsets tested over the same space for a trajectory starting from point $\left(x_1, x_2\right)$.\\
The observer on the left side was trained on the gaussian distributed initial conditions $\mathcal{X}$ where $\mathcal{X}_1 = \left[-1, 1\right] \times \left[-1,1\right]$. Due to the boundedness property of reverse duffing oscillator, the observer did not see the data in the outer range during training, so the estimation error is large, while the right side observer was trained on $\mathcal{X}_2 = \left[-3, 3\right] \times \left[-3, 3\right]$ and has learned the data in this range during the training process and has a significantly smaller error. Figure \ref{compare_x12} illustrated the estimated state trajectories under the initial condition $x(t_0)=(2,0)$ which is outside the training domain of the first observer while inside that of the second observer. Unsurprisingly, the results indicate the extremely limited generalization capability which is consistent with the results in \cite{numericalKKL}.  The PDE constraint \eqref{PDE} can also be added to the loss function to constrain the learning of $T$ as \cite{PDE-KKL} inspired to improve the generalization, but since the effect is not quite significant, the results are not shown here.
\begin{wrapfigure}{r}{7cm}
  \centering
  \includegraphics[scale=0.45]{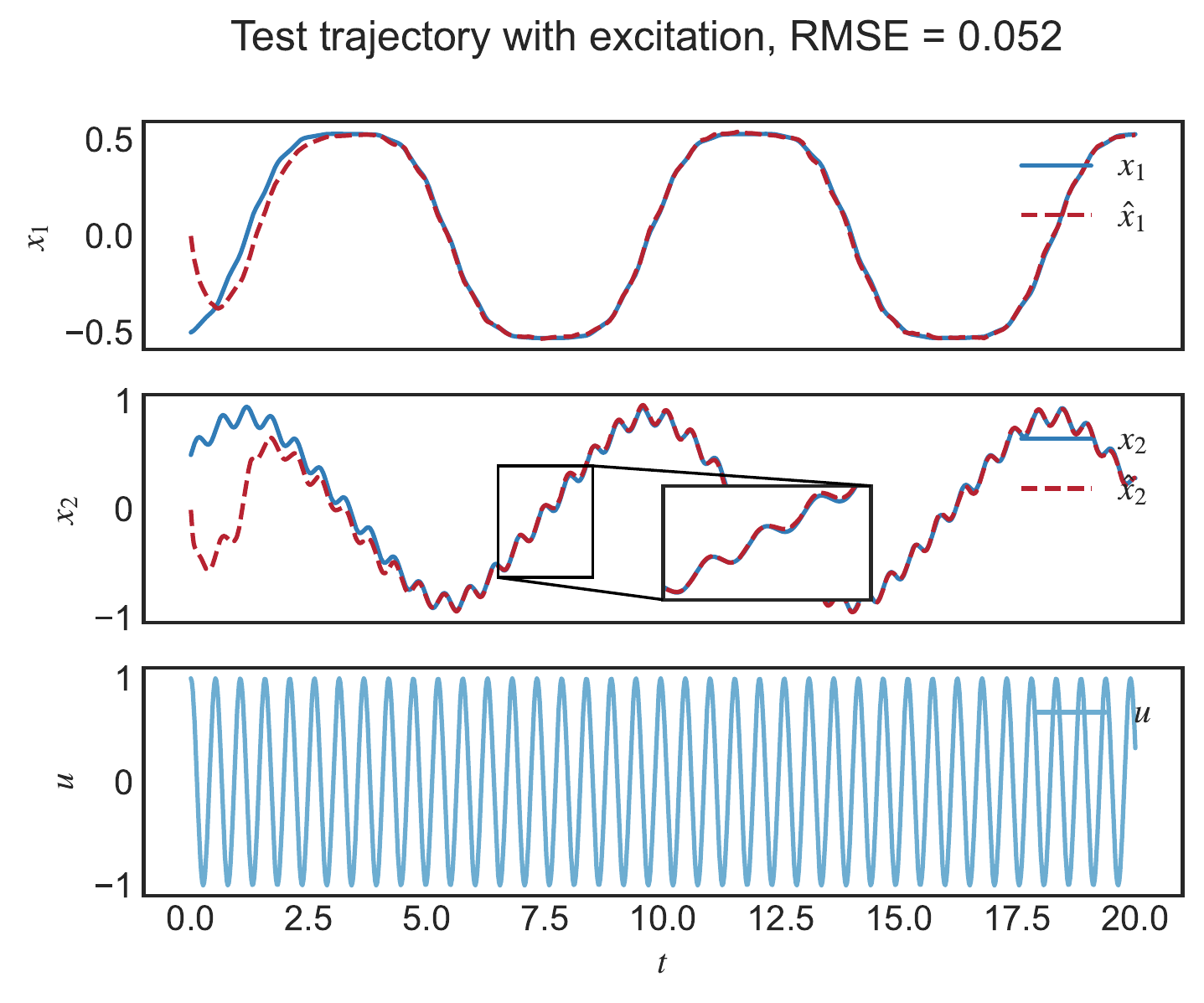}
  \vspace{-20pt}
  \caption{Test trajectory of oscillator with excitation}
  \label{duff-exc}
\end{wrapfigure}
\begin{remark} 
As introduced in \cite{kkl-non}, a stationary transformation $\mathcal{T}$ and $\mathcal{T}^*$ for an autonomous system can be used for time-varying systems
\begin{equation}
\dot x = f(x)+g(x)u
\end{equation}
where the latent state $z$ satisfies
\begin{equation}
\dot z = Dz + Fy + \varphi(z)u,\ \quad \varphi(z) = \frac{\partial \mathcal{T}}{\partial x}(\mathcal{T}^*(z))g(\mathcal{T}^*(z))
\end{equation}
In this case, mapping $\mathcal{T}$ is also needed, hence the loss function for training autonomous system observer is modified as $\ell = \int_{t_0}^{t_f} \lvert z\left(t\right) - \mathcal{T}\left(x\left(t\right), \eta_1\right)  \rvert^2 + \lvert x\left(t\right) - \mathcal{T^*}\left(\mathcal{T}\left(x\left(t\right), \eta_1\right), \eta_2\right) \rvert^2 dt$. Finally, the observer learned by our approach can be easily implemented to a non-autonomous system with excitation. When the system \eqref{oscillator} has an excitation as $\dot x_2\left(t\right)=-x_1\left(t\right) + u\left(t\right)$, $u\left(t\right) = \cos (12*t)$, the previously learned observer is shown in Figure \ref{duff-exc} which still exhibits satisfactory performance. 
\end{remark}
\section{Conclusions}
\label{5}
In this paper, we propose a Neural ODEs-based framework to design nonlinear state observers and cast the design problem into an optimal control problem. In particular, for KKL observer, we discuss the relationship between the observer design and its trade-off between convergence speed and robustness, and then use Neural ODEs to learn more robust observers than those previously addressed. Besides, our framework can then be extended to non-autonomous systems by applying stationary transformation. We have evaluated our approach through numerical simulations and the results show that our method outperforms the existing baseline.\\
There are also a number of outstanding problems to be addressed in the future. One of the most notable problem is that current learning-based KKL observers all suffer from poor generalization ability, and theoretical guarantees of a certain level of generalization remains an open problem. The application of observer methods to output prediction and system identification problems is also a direction of interest.

\bibliography{reference}{}

\section{Appendix}
\subsection{Proofs and Additional Theoretical Results}
\subsubsection{Proof of Proposition \ref{pro1}}
\label{profor3}
\begin{proof}
Define the augmented objective function with a Lagrange multiplier $p$ as
\begin{equation}
\mathcal{L} = \ell - \int_{t_0}^{t_f} p(t)\left[\dot z(t) - \mathcal{F}\left(z(t), y(t), \theta\right)\right] dt
\label{aug}
\end{equation}
and $\dot z - \mathcal{F} =0$ always holds by construction, so $p(t)$ can be freely assigned while $\frac{d \mathcal{L}}{d \theta} = \frac{d \ell}{d \theta}$. As to the integration part on right hand side of \eqref{aug}, we have
$$
\begin{aligned}
\int_{t_0}^{t_f}p(t)\left(\dot z - \mathcal{F}\right)dt & = p(t)z(t)\big{|}_{t_0}^{t_f} - \int_{t_0}^{t_f}\dot p(t)z(t)dt - \int_{t_0}^{t_f}p(t)\mathcal{F}dt\\
& = p\left(t_f\right)z\left(t_f\right) - p\left(t_0\right)z\left(t_0\right) - \int_{t_0}^{t_f}\left(\dot p(t)z(t) + p(t)\mathcal{F}\right)dt
\end{aligned}
$$
Hence,
$$
\mathcal{L} = - p\left(t_f\right)z\left(t_f\right) + p\left(t_0\right)z\left(t_0\right) + \int_{t_0}^{t_f}\left(\dot p(t)z(t) + p(t)\mathcal{F} + L \right)dt
$$
Then the gradient of $\ell$ with respect to $\theta$ can be computed as
\begin{equation}
\frac{d \ell}{d \theta} = \frac{d \mathcal{L}}{d \theta} = -p(t_f)\frac{d z(t_f)}{d \theta} + \int_{t_0}^{t_f}\left(\dot p(t)\frac{d z(t)}{d \theta} + p(t)\left(\frac{\partial \mathcal{F}}{\partial \theta} + \frac{\partial\mathcal{F}}{\partial z}\frac{d z}{d \theta}\right) + \frac{\partial L}{\partial \theta} + \frac{\partial L}{\partial z}\frac{d z}{d \theta}\right)dt
\label{gradl}
\end{equation}
Now if we set this Lagrange multiplier as the adjoint state for the Hamiltonian function
\begin{equation}
H\left(z,p,\theta\right) = p\mathcal{F} + L
\end{equation}
and according to PMP, the optimality conditions requires 
\begin{equation}
\dot z = \frac{\partial H}{\partial p} = \mathcal{F},\ \dot p = -\frac{\partial H}{\partial z} = -p\frac{\partial \mathcal F}{\partial z} - \frac{\partial L}{\partial z}
\label{PMP-condition}
\end{equation}
with initial conditions $z(t_0) = z_0$ and $p(t_f)=\frac{\partial \Phi}{\partial z(t_f)} = \vmathbb{0}_{n_{z}}$.
Substituting \eqref{PMP-condition} into \eqref{grad}, it can be obtained that
\begin{equation}
\frac{d \ell}{d \theta} = \int_{t_0}^{t_f}\left(p(t)\frac{\partial \mathcal{F}}{\partial \theta} + \frac{\partial L}{\partial \theta}\right)dt = \int_{t_f}^{t_0}\left(- p(t)\frac{\partial \mathcal{F}}{\partial \theta} - \frac{\partial L}{\partial \theta}\right)dt = \int_{t_f}^{t_0} -\frac{\partial H}{\partial \theta}dt
\end{equation}
proving the result.
\end{proof}
\subsubsection{Proof of Proposition \ref{robust}}
\label{profor4}
\begin{proof}
For system \eqref{noise} and \eqref{KKL-z}, and $D$ is a diagonal matrix whose eigenvalues are $(\lambda_1,...\lambda_{d_z})$,
$$
\dot {\overbrace{z\left(t\right) - \mathcal{T}\left(x\left(t\right)\right)}} = \dot z(t) - \frac{\partial \mathcal{T}}{\partial x}(x)\left(f(x(t)) + w(t)\right)
$$
where $\mathcal {T}$ is obtained by PDE \eqref{PDE}, then
$$
\begin{aligned}
\dot {\overbrace{z\left(t\right) - \mathcal{T}\left(x\left(t\right)\right)}} & = Dz(t) + F(h(x) + v(t)) - \frac {\partial \mathcal{T}}{\partial x}\left(x\right) w\left(t\right) - D\mathcal{T}(x(t))-Fh(x)\\
& = D\left(z - \mathcal{T}\left(x\left(t\right)\right)\right) + Fv\left(t\right) - \frac {\partial \mathcal{T}}{\partial x}\left(x\right) w\left(t\right)
\end{aligned}
$$
Then it can be derived that
$$
z\left(t\right) - \mathcal{T}\left(x\left(t\right)\right)  = \exp\left(Dt\right)\left(z\left(t_0\right) - \mathcal{T}\left(x\left(t_0\right)\right)\right) + \int_{t_0}^{t}\exp\left(D\left(t - s\right)\right)\left(Fv\left(s\right) - \frac {\partial \mathcal{T}}{\partial x}\left(x\right) w\left(s\right)\right)ds
$$
Hence,
$$
\begin{aligned}
\lvert  z\left(t\right) - \mathcal{T}\left(x\left(t\right)\right) \rvert & \leq \exp\left(\max_i\{\Re\lambda_i\}t\right)\lvert z\left(t_0\right) - \mathcal{T}\left(x\left(t_0\right)\right)\rvert \\
& + \int_{t_0}^{t}\exp\left(\max_i\{\Re\lambda_i\}\left(t - s\right)\right)\left(\lvert Fv\left(s\right) \rvert + \lvert \frac {\partial \mathcal{T}}{\partial x}\left(x\right) w\left(s\right)\rvert \right)ds\\
& \leq \exp\left(\max_i\{\Re\lambda_i\}t\right)\lvert z\left(t_0\right) - \mathcal{T}\left(x\left(t_0\right)\right)\rvert\\
& + \int_{t_0}^{t}\exp\left(\max_i\{\Re\lambda_i\}\left(t - s\right)\right)ds \times \left(
\bar v + L_{\mathcal{T}_x}\bar w\right)\\
& = \exp\left(\max_i\{\Re\lambda_i\}t\right)\lvert z\left(t_0\right) - \mathcal{T}\left(x\left(t_0\right)\right)\rvert\\
& + \frac{1 -\exp\left(\max_i\{\Re\lambda_i\}\left(t - t_0\right)\right)}{\lvert \max_i\{\Re\lambda_i\} \rvert} \times \left(
\bar v + L_{\mathcal{T}_x}\bar w\right)\\
& \leq \exp\left(\max_i\{\Re\lambda_i\}t\right)\lvert z\left(t_0\right) - \mathcal{T}\left(x\left(t_0\right)\right)\rvert + \frac{L_{\mathcal{T}_x}\bar w + \bar v}{\lvert \max_i\{\Re\lambda_i\} \rvert}
\end{aligned}
$$
And based on the assumption that $\lvert x_1 - x_2 \rvert \leq L_\mathcal{T} \lvert \mathcal{T}\left(x_1\right) - \mathcal{T}\left(x_2\right) \rvert$ indicating the uniformly injective of $\mathcal{T}$, (which was proved in [\cite{convergence}, Lemma 3.2] that for almost all Hurwitz diagonal matrices $D$, it holds under an extra observability assumption) then for all $(z,x)\in \mathbb{R}^{n_z} \times \mathcal {X}$, it yields \citep{convergence}
\begin{equation}
\lvert \mathcal{T}^*\left(z\right) - x\rvert \leq \sqrt n_x L_\mathcal{T} \lvert z - \mathcal{T}\left(x\right)\rvert
\label{root}
\end{equation}
Hence,
\begin{equation}
\begin{aligned}
\lvert x\left(t\right) - \tilde x\left(t\right) \rvert & = \lvert x\left(t\right) - \mathcal{T}^*\left(z\left(t\right)\right)\rvert\\
& \leq \sqrt n_x L_\mathcal{T} \exp\left(\lambda_{max}t\right)\lvert z\left(t_0\right) - \mathcal{T}\left(x\left(t_0\right)\right)\rvert + \frac{\sqrt n_x L_\mathcal{T} }{\lvert \lambda_{max}\rvert}\left(L_{\mathcal{T}_x}\bar w + \bar v\right)
\end{aligned}
\end{equation}
When the system satisfies complete observability \cite{KKL2}, and the mapping is defined with $D$ whose eigenvalues are $k\lambda_i$ that $k$ is sufficiently large, the corresponding mapping is defined as $\mathcal{T}_k$ and $\mathcal{T}_k^*$.\\
$\mathcal{T}_k$ is the solution to PDE \eqref{PDE} where $D$ is a diagonal Hurwitz matrix, then as derived in \cite{convergence} Section 5.3, we can follow the proof that
$$
\begin{aligned}
\frac{\partial \mathcal{T}_k}{\partial x}(x) &= \int_{-\infty}^{0}\exp(-kDt)\frac{\partial{h}}{\partial x}\left(\breve x(x,t)\right)\frac{\partial \breve x}{\partial x}(x,t)dt\\
& \leq \int_{-\infty}^{0}\exp\left({-k\max_i\{\Re\lambda_i\}t}\right) \sqrt{n_x}\exp(\rho t)dt \times \max_x \lvert \frac{\partial h}{\partial x}(x)\rvert\\
& = \frac{\sqrt{n_x}}{\lvert k \max_i\{\Re\lambda_i\} \rvert} \max_x \lvert \frac{\partial h}{\partial x}(x)\rvert
\end{aligned}
$$
where $\breve x$ is the solution of the modified dynamical system as stated in \cite{convergence} Section 3.2, and $\rho$ is a negative number and $\lambda_i \leq \rho$ defined in Section 5.3.\\
Compared with $\lvert \frac{\partial \mathcal{T}}{\partial x}\left(x\right)\rvert \leq L_{\mathcal{T}_x}$ which can be seen as the case that $k=1$, it can be checked that
\begin{equation}
\lvert \frac{\partial \mathcal{T}_k}{\partial x}\left(x\right)\rvert \leq \frac{L_{\mathcal{T}_x}}{k}
\end{equation} 
where $L_{\mathcal{T}_x}$ is the upper bound of $\lvert \frac{\partial \mathcal{T}}{\partial x}\left(x\right)\rvert$ (or can be interpreted as $\lvert \frac{\partial \mathcal{T}_1}{\partial x}\left(x\right)\rvert$ for the case that $k=1$) as the assumption stated in Proposition \ref{robust}. Combined with steps above, it yields
\begin{equation}
\lvert  z\left(t\right) - \mathcal{T}_k\left(x\left(t\right)\right) \rvert \leq \exp\left(k \max_i\{\Re\lambda_i\}t\right)\lvert z\left(t_0\right) - \mathcal{T}_k\left(x\left(t_0\right)\right)\rvert + \frac{1}{\lvert k \max_i\{\Re\lambda_i\} \rvert}\left(\frac{L_{\mathcal{T}_x}}{k}\bar w + \bar v\right)
\label{z-bound}
\end{equation}
Based on [\cite{convergence}, Theorem 4.1], it can be derived that for $k\geq 1$
\begin{equation}
\lvert x_1 - x_2 \rvert \leq k^{n_z} L_\mathcal{T} \lvert \mathcal{T}_k\left(x_1\right) - \mathcal{T}_k\left(x_2\right) \rvert
\end{equation}
and as previously stated, it yields
\begin{equation}
\lvert \mathcal{T}_k^*\left(z\right) - x\rvert \leq k^{n_z} \sqrt n_x L_\mathcal{T} \lvert z - \mathcal{T}_k\left(x\right)\rvert
\end{equation}
Hence, the proposition holds that
$$
\lvert x\left(t\right) - \tilde x\left(t\right) \rvert \leq k^{n_z} \sqrt n_x L_\mathcal{T} \exp\left(k \lambda_{max}t\right)\lvert z\left(t_0\right) - \mathcal{T}_k\left(x\left(t_0\right)\right)\rvert + \frac{k^{n_z}\sqrt n_x L_\mathcal{T} }{\lvert k \lambda_{max}\rvert}\left(\frac{L_{\mathcal{T}_x}}{k}\bar w + \bar v\right)
$$
\end{proof}
\subsubsection{Proof of Learning-based KKL Observer Robustness to Approximation Error}
\label{proforrem}
\begin{proof}
Let $z^*\left(t\right) = \mathcal{T}\left(x\left(t\right)\right)$ (i.e., the solution of $\dot z = Dz + Fy$ at time $t$ under the initial condition $z\left(t_0\right) = \mathcal{T}\left(x\left(t_0\right)\right)$) and $z\left(t\right)$ is the solution at time $t$ from any arbitrary initial state, then 
$$
\begin{aligned}
\hat x\left(t\right) - x\left(t\right) &= \hat {\mathcal{T}}^*\left(z\left(t\right)\right) - \mathcal{T}^*\left(z^*\left(t\right)\right)\\
& = \hat {\mathcal{T}}^*\left(z\left(t\right)\right) - \hat{\mathcal{T}}^*\left(z^*\left(t\right)\right) - \varepsilon\left(t\right)
\end{aligned}
$$
Since the activation functions of a neural network are Lipschitz continuous, then $\hat{\mathcal{T}}^*$ is also Lipschitz, which means that there exists $l$ such that for each $z$ and $z^*$, $\lvert \hat{\mathcal{T}}^*\left(z\left(t\right)\right) - \hat{\mathcal{T}}^*\left(z^*\left(t\right)\right) \rvert \leq l \lvert z(t) - z^*(t)\rvert$ holds. Hence, it can be derived that
$$
\begin{aligned}
\lvert \hat x\left(t\right) - x\left(t\right) \rvert &\leq \lvert \hat{\mathcal{T}}^*\left(z\left(t\right)\right) - \hat{\mathcal{T}}^*\left(z^*\left(t\right)\right) \rvert + \lvert \varepsilon\left(t\right) \rvert\\
& \leq l \lvert z\left(t\right) - \mathcal T \left(x\left(t\right)\right)\rvert + \bar \varepsilon\
\end{aligned}
$$
Followed by \eqref{z-bound}, the proof completes.
\end{proof}
\subsection{Experimental Details}
\subsubsection{Example 1}
\paragraph{Network Structure}
In this task, the network for $\hat g$ is a three-layer feedforward neural network and hidden layer is of 16 units.
\paragraph{Training Details}
The training data (initial state) is generated from a uniform random distribution $\mathcal{X} = \left[-5, 5\right] \times \left[-5,5\right]$, and the dynamics are solved over the time interval $\left[0, t_f\right]$ with $t_f = 50s$. The ODE solver for Neural ODEs is chosen as RK4 with step size 0.02s. The training has been carried out for 1000 epoches using GD with learning rate $10^{-3}$, batch size 50.
\subsubsection{Example 2}
\begin{figure}[ht]
  \centering
  \includegraphics[scale=0.35]{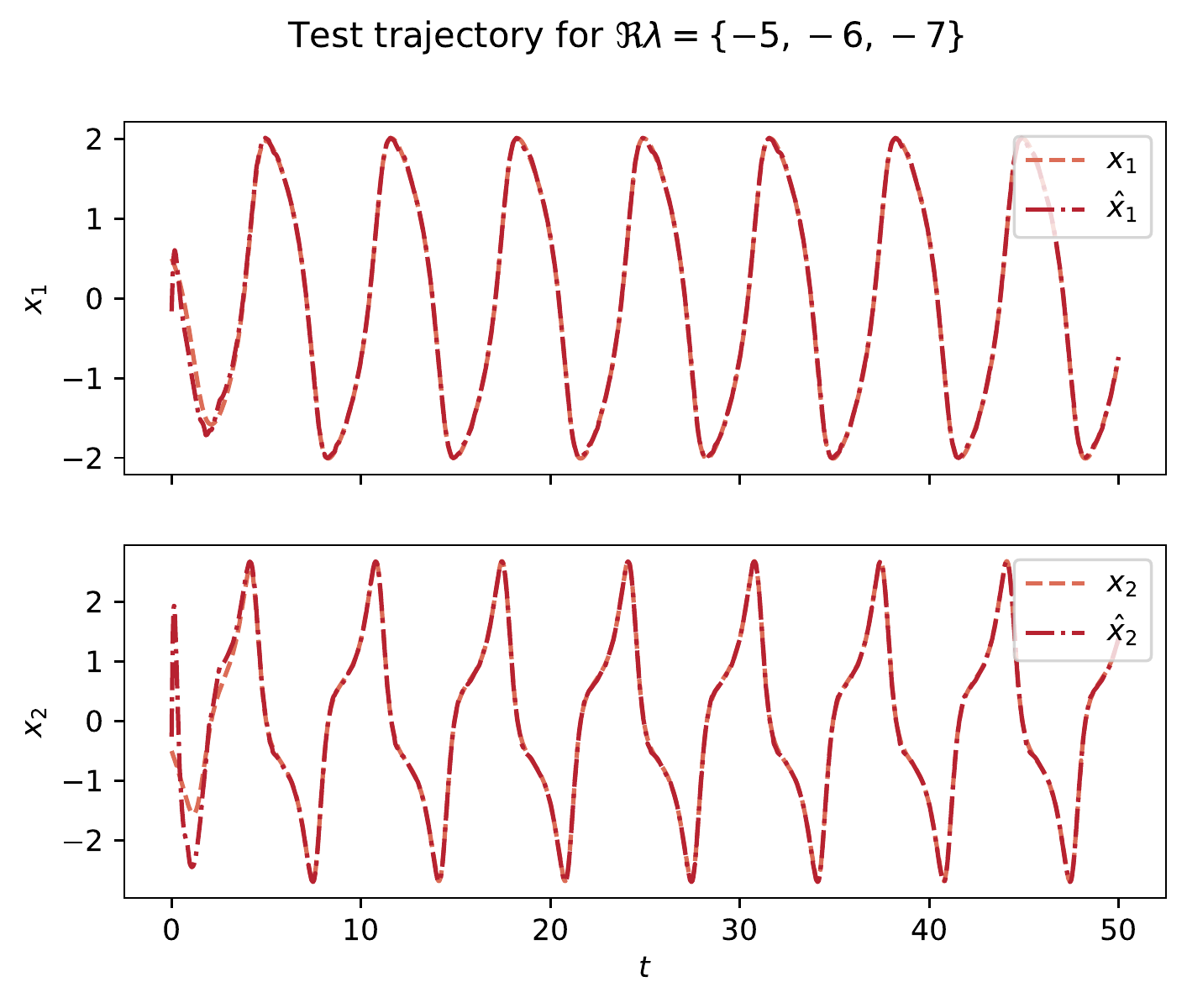}
  \hspace{-0.2cm}
  \includegraphics[scale=0.35]{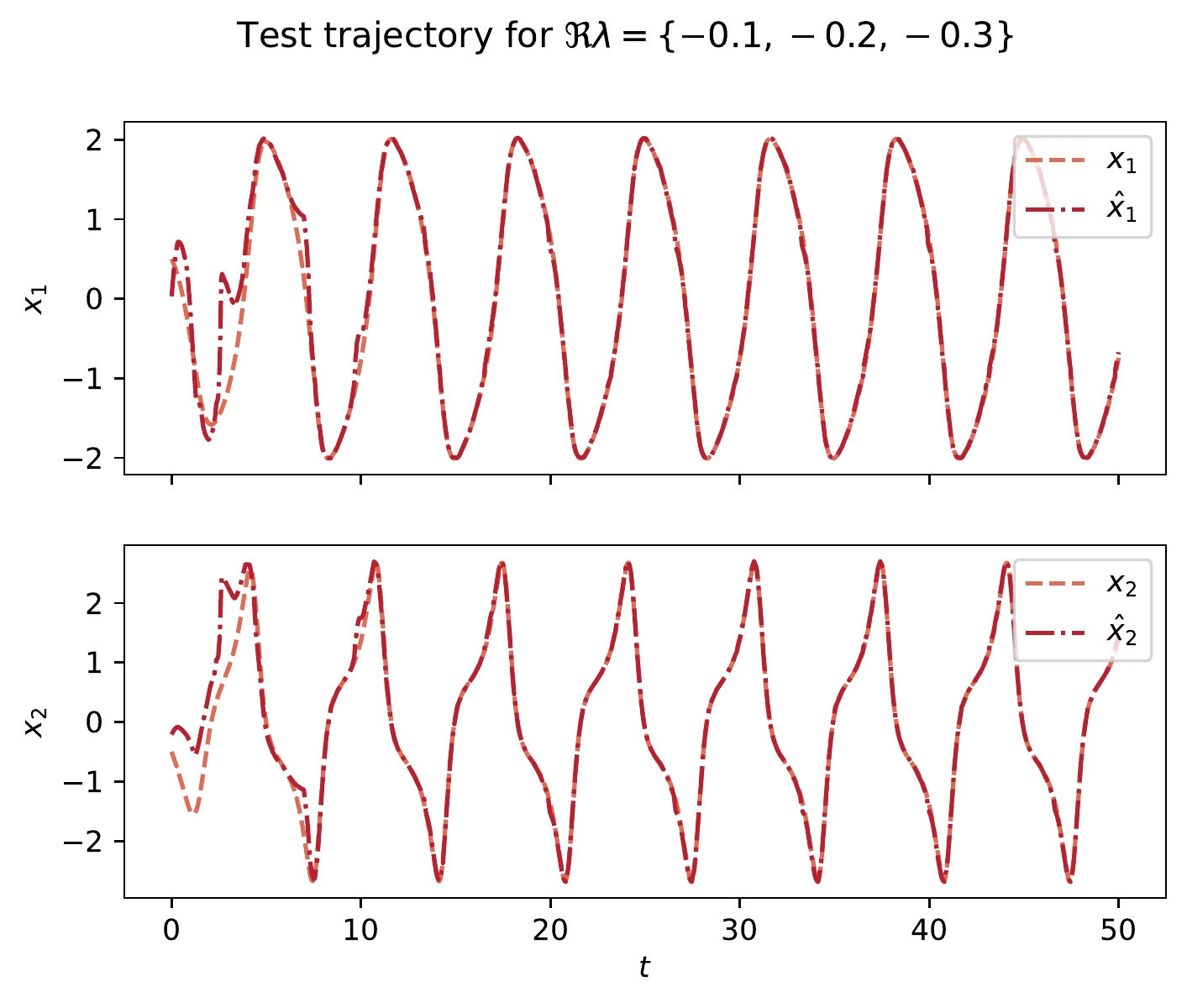}
  \hspace{-0.2cm}
  \includegraphics[scale=0.35]{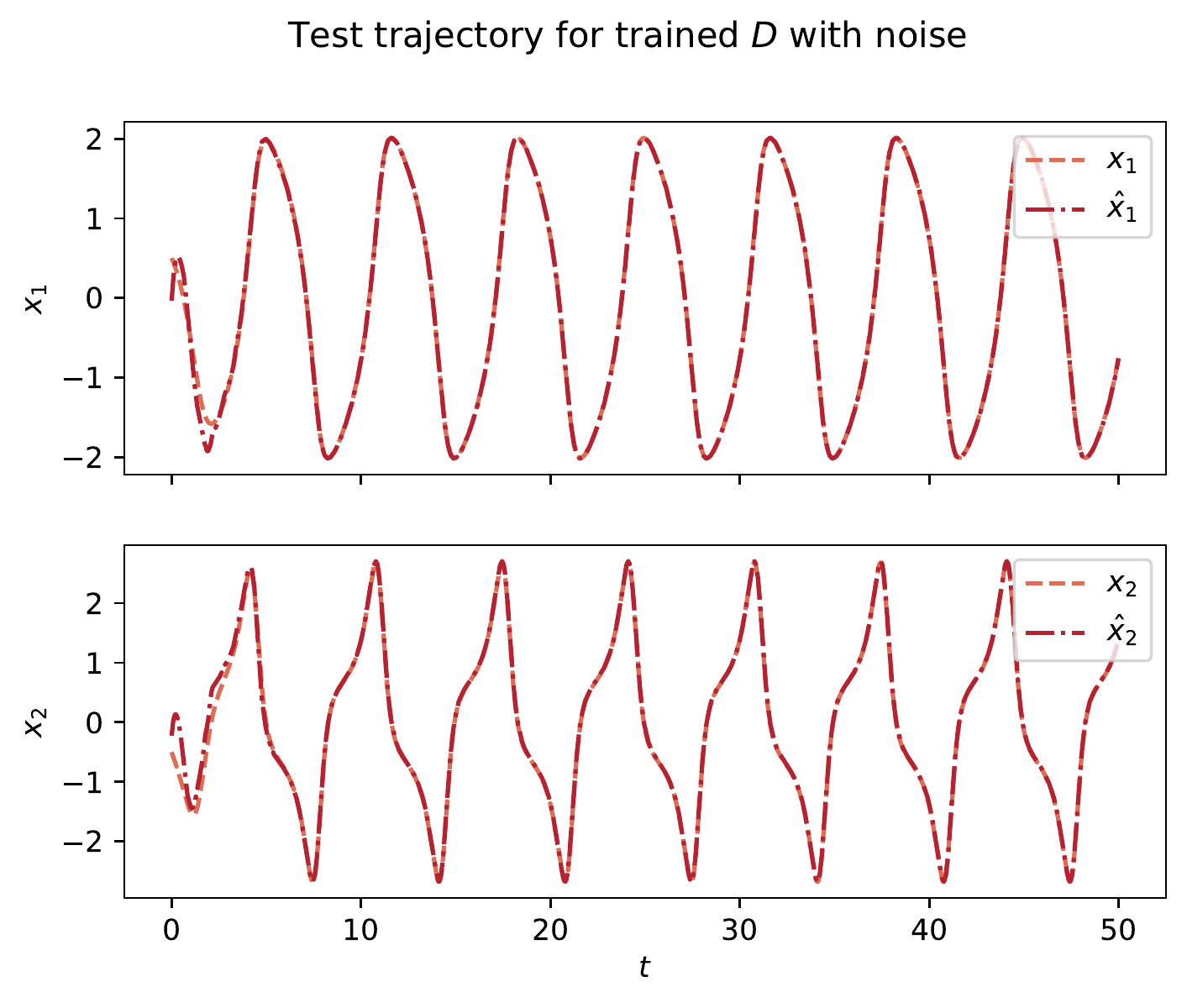}
  \caption{Test trajectories of Van der Pol for $x_0 = (-0.5, 0.5)$ without measurement noise}
  \label{vanderpol-compare-without}
\end{figure}

\begin{figure}[ht]
  \centering
  \includegraphics[scale=0.45]{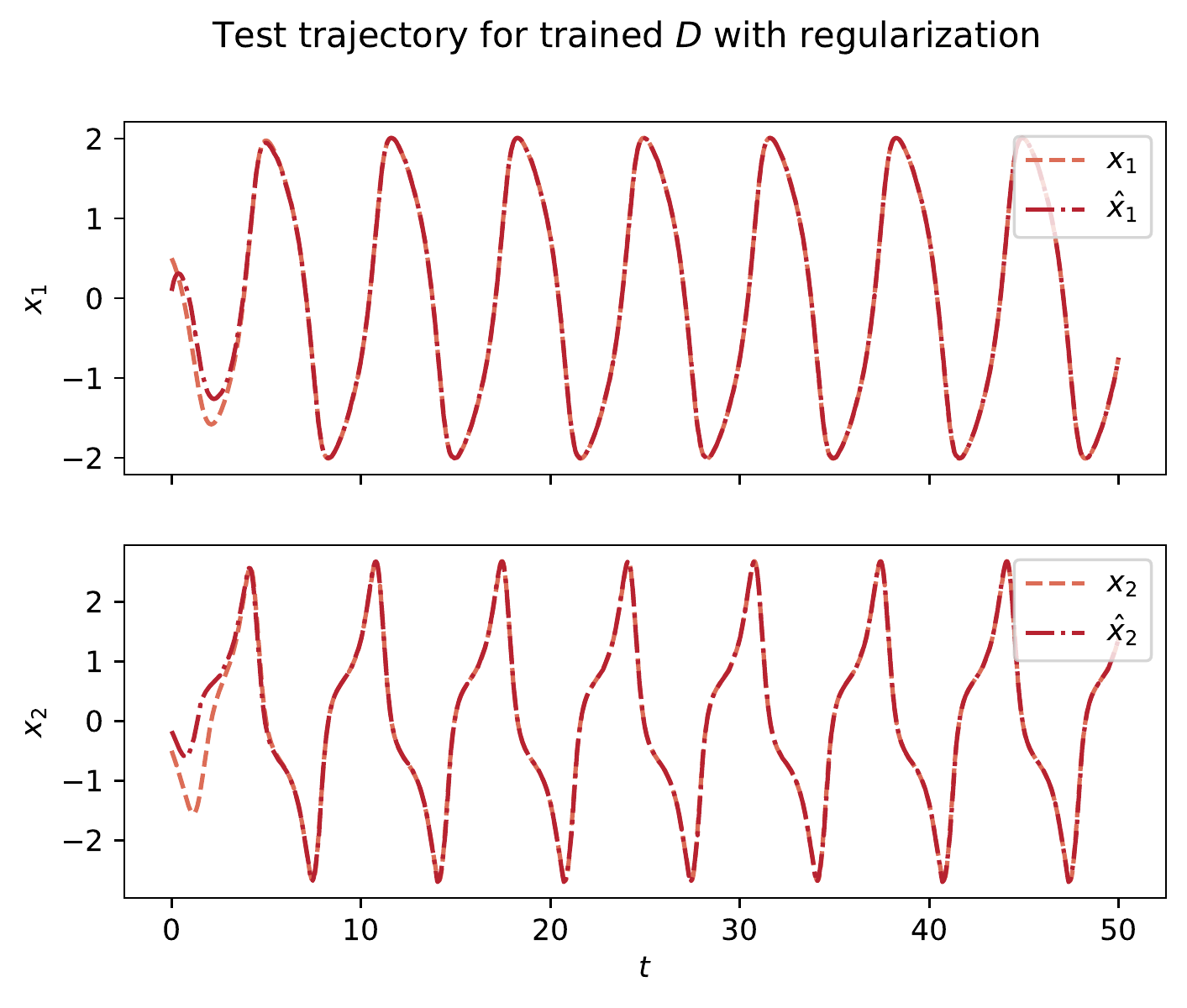}
  \hspace{-0.2cm}
  \includegraphics[scale=0.45]{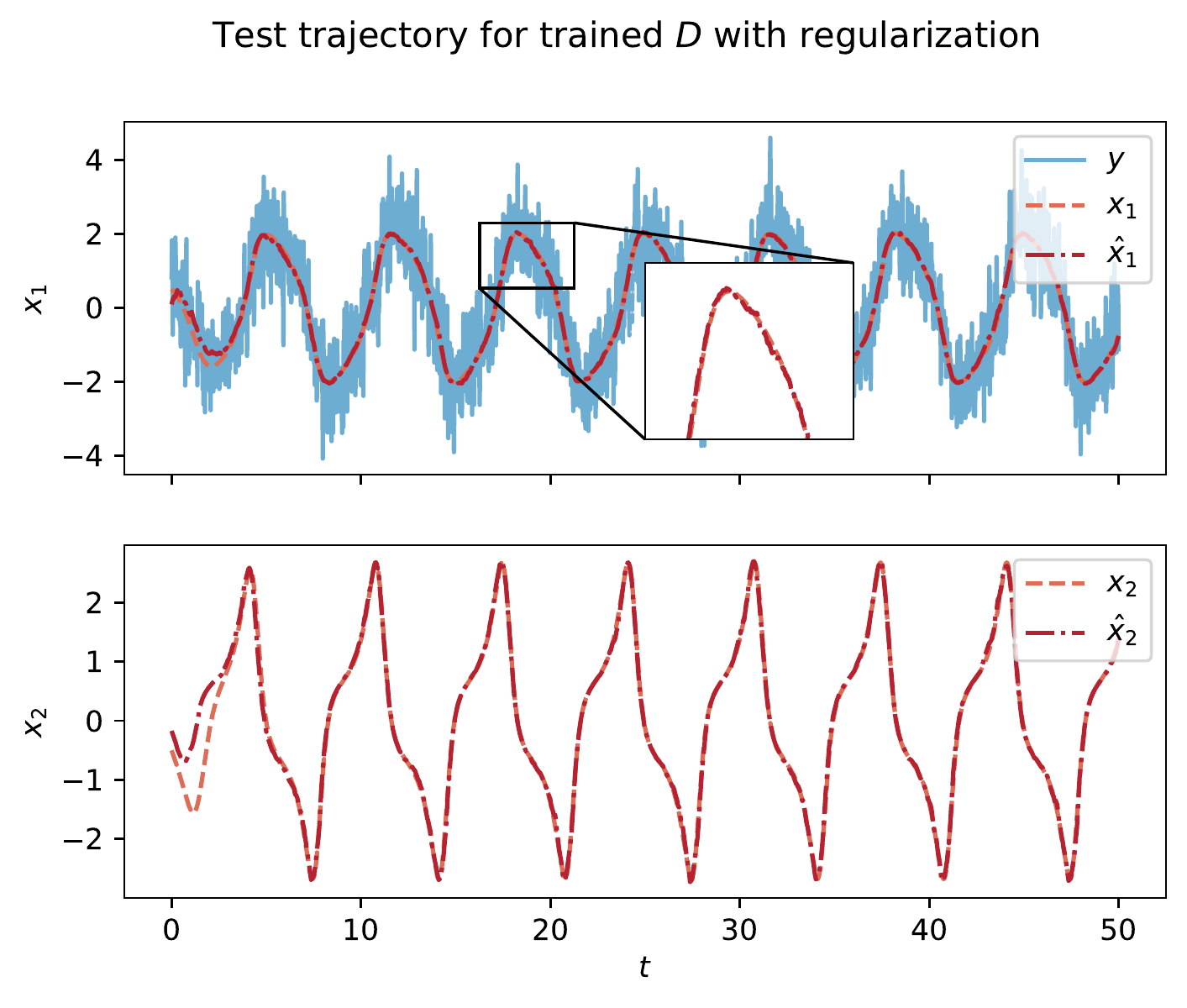}
  \caption{Test trajectories of Van der Pol for $x_0 = (-0.5, 0.5)$ using regularization technique}
  \label{vanderpol-compare-reg}
\end{figure}

\begin{figure}[ht]
\centering
\includegraphics[scale=0.6]{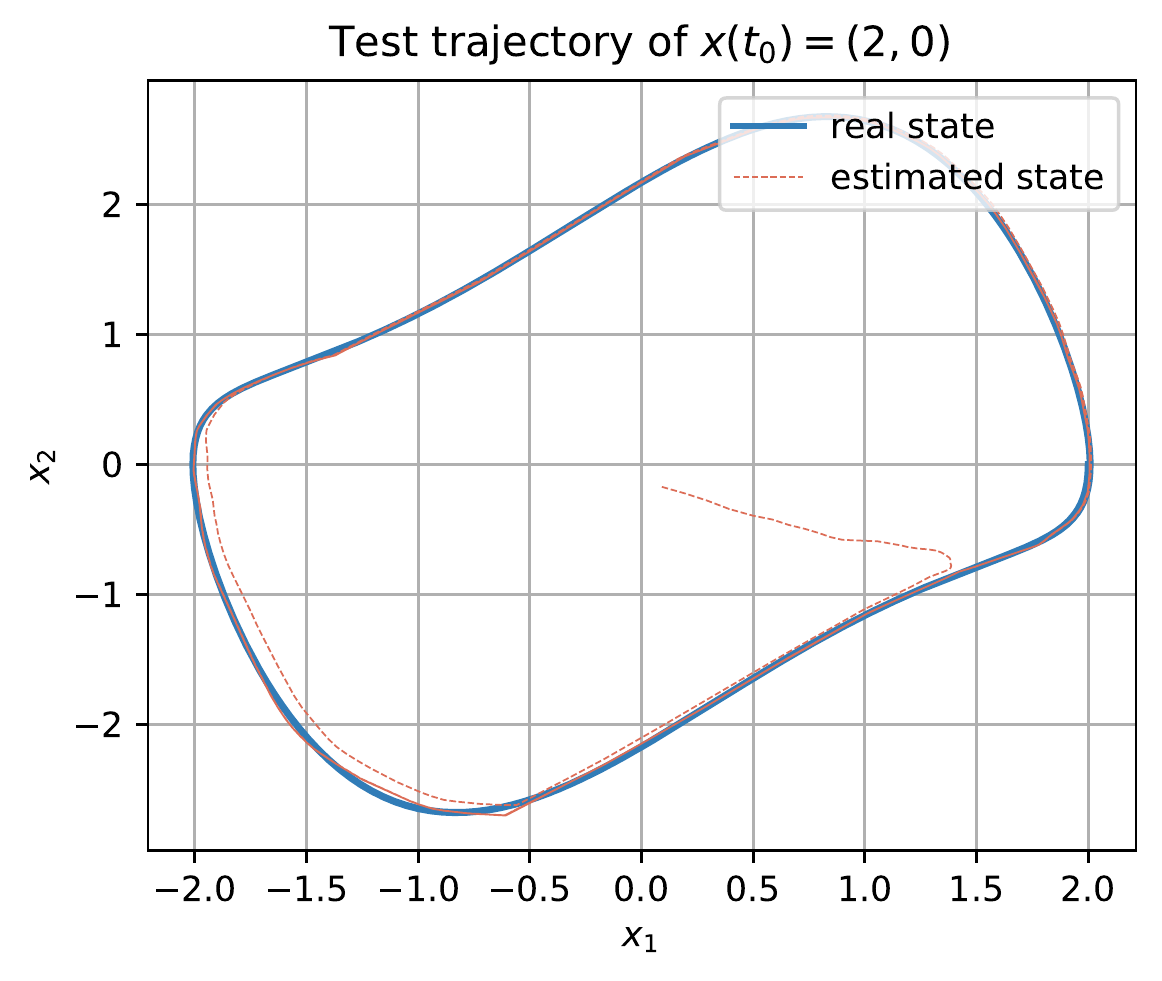}
\caption{Estimated trajectory of observer trained with regularization technique for $x(t_0) = (2, 0)$}
\label{reg-traj}
\end{figure}

\begin{figure}[ht]
  \centering
  \includegraphics[scale=0.45]{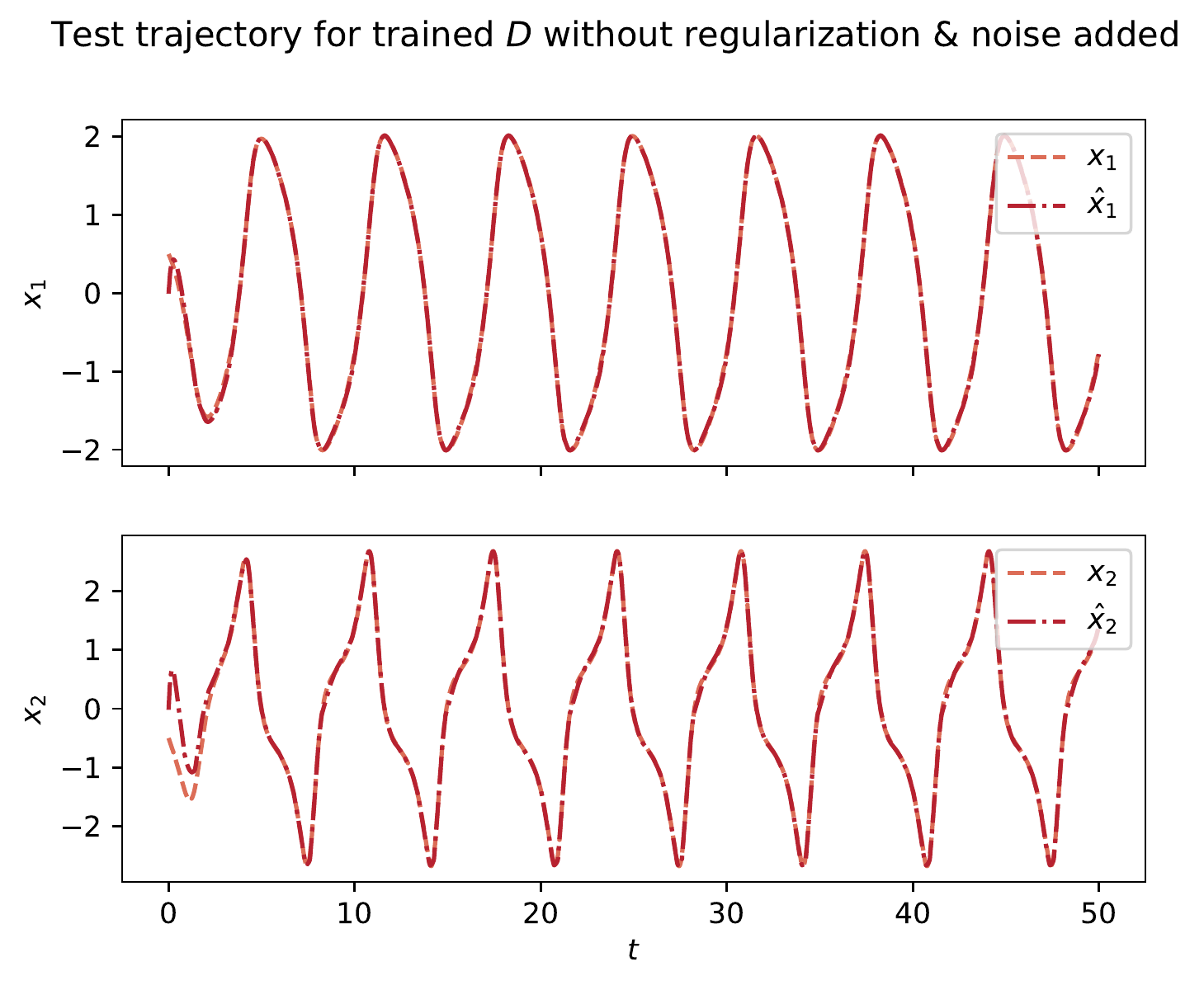}
  \hspace{-0.2cm}
  \includegraphics[scale=0.45]{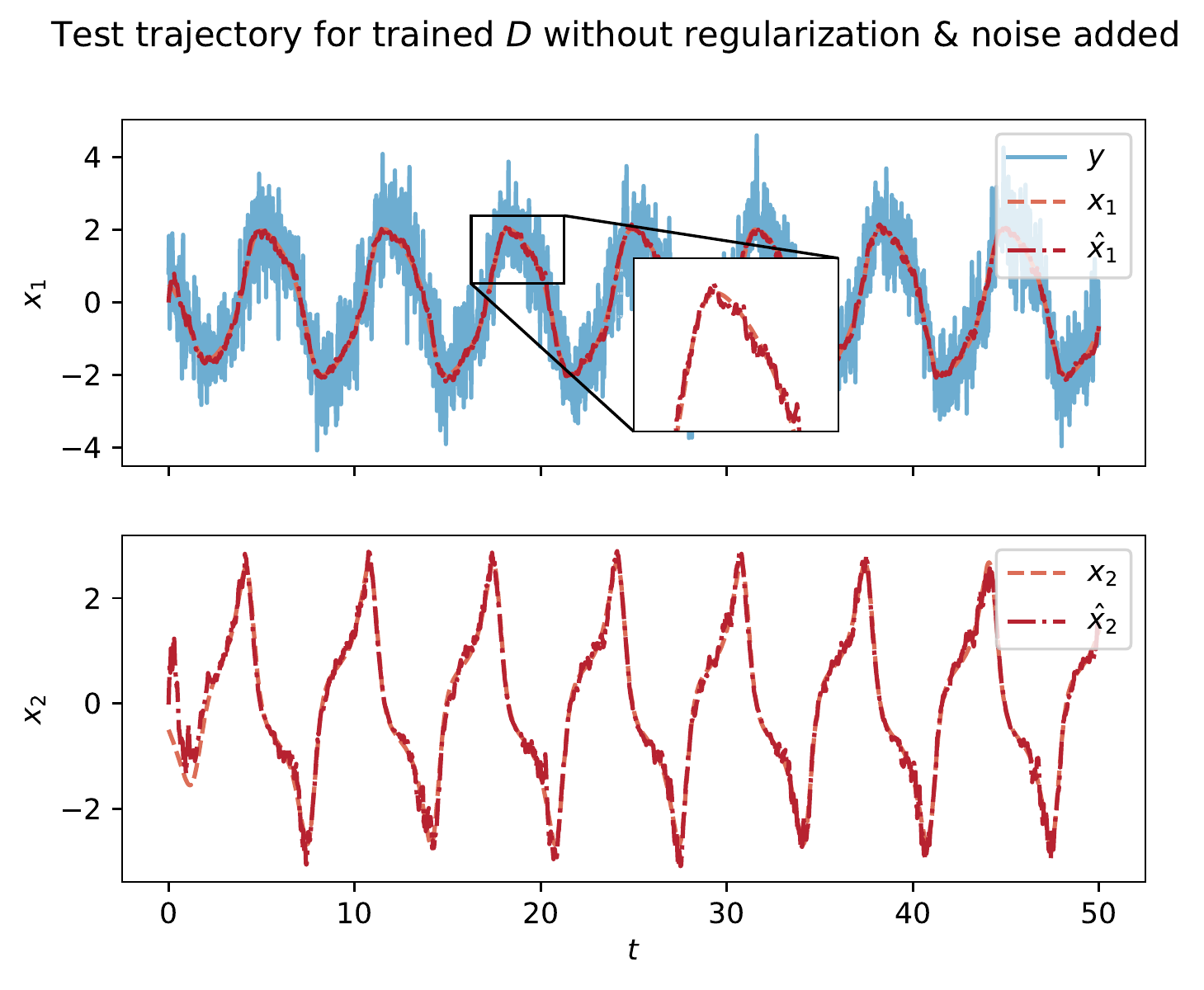}
  \caption{Test trajectories of Van der Pol for $x_0 = (-0.5, 0.5)$ without robustness improvement skills}
  \label{vanderpol-compare-no}
\end{figure}

\paragraph{Network Structure}
In this task, the network for $\mathcal{T}^*$ is multilayer perceptron with four hidden layers of 50 neurons each. $D$ is a linear layer.
\paragraph{Training Details}
The training data (initial state) is generated from a uniform random distribution $\mathcal{X} = \left[-1, 1\right] \times \left[-1,1\right]$, and the dynamics are solved over the time interval $\left[0, t_f\right]$ with $t_f = 50s$. The ODE solver for Neural ODEs is chosen as RK4 with step size 0.02s. The training has been carried out for 1000 epoches using Adam optimizer with learning rate $10^{-3}$ and ExponentialLR with weight decay $10^{-4}$, batch size 50.
\paragraph{Additional Results}
The test results that without noise in measurement for the three cases in Figure \ref{vanderpol-compare} are shown in Figure \ref{vanderpol-compare-without}, 
and the eigenvalues of learned $D$ (with noise added in training) are $\lambda = \{-0.81, -0.83, -0.63\}$.\\
As to the case that training using regularization technique, the eigenvalues of learned $D$ are $\lambda = \{-0.66, -0.99, -1.77\}$ and the test results with and without noise are shown in Figure \ref{vanderpol-compare-reg}. Figure \ref{reg-traj} illustrates the estimated trajectory starting with $x(t_0) = (2, 0)$.\\
\\
If no noise added in measurement  during training and regularization technique is not used, but $D$ is still learned via Neural ODEs, the loss function is as \eqref{loss} and the test results are shown in Figure \ref{vanderpol-compare-no}.\\
\\
We should point out that the first two columns in Table \ref{vanderpol-rmse} are the results obtained using the method proposed by \cite{numericalKKL}, so this is the quantitative result of comparing our method with the previous ones. 
\begin{table}[t]
\large
\vspace{-10pt}
\caption{RMSE of designed Observers tested in different scenarios}
\label{vanderpol-rmse-add}
\resizebox{.95\columnwidth}{!}{
\begin{tabular}{c|c|c|c}
\hline
\diagbox{Test Scenario}{Eigenvalues of $D$} & \{-5, -6, -7\} & \{-0.1, -6, -7\} & \{-0.1, -0.2, -0.3\} \\ \hline
No Noise       &$\mathbf {0.0548}$ &$0.1786$ &$0.2080$  \\ \hline
Gaussian Noise $\mathcal{N}(0,0.5)$ &$0.1160$  &$0.1903$  &$0.2273$ \\ \hline
Uniform Noise $\mathcal{U}(-3,3)$              &$0.3205$  &$0.2586$  &$0.2560$ \\ \hline
\end{tabular}}
\vspace{-10pt}
\end{table}
\\
Since the largest eigenvalue plays a key role in our proof (although in the process we are multiplying all the eigenvalues by a factor $k$). So in Table \ref{vanderpol-rmse-add} we have added a comparison of the results of changing only the largest eigenvalue, in line with our conclusion.
\subsubsection{Example 3}
\begin{figure}[ht]
  \centering
  \includegraphics[scale=0.45]{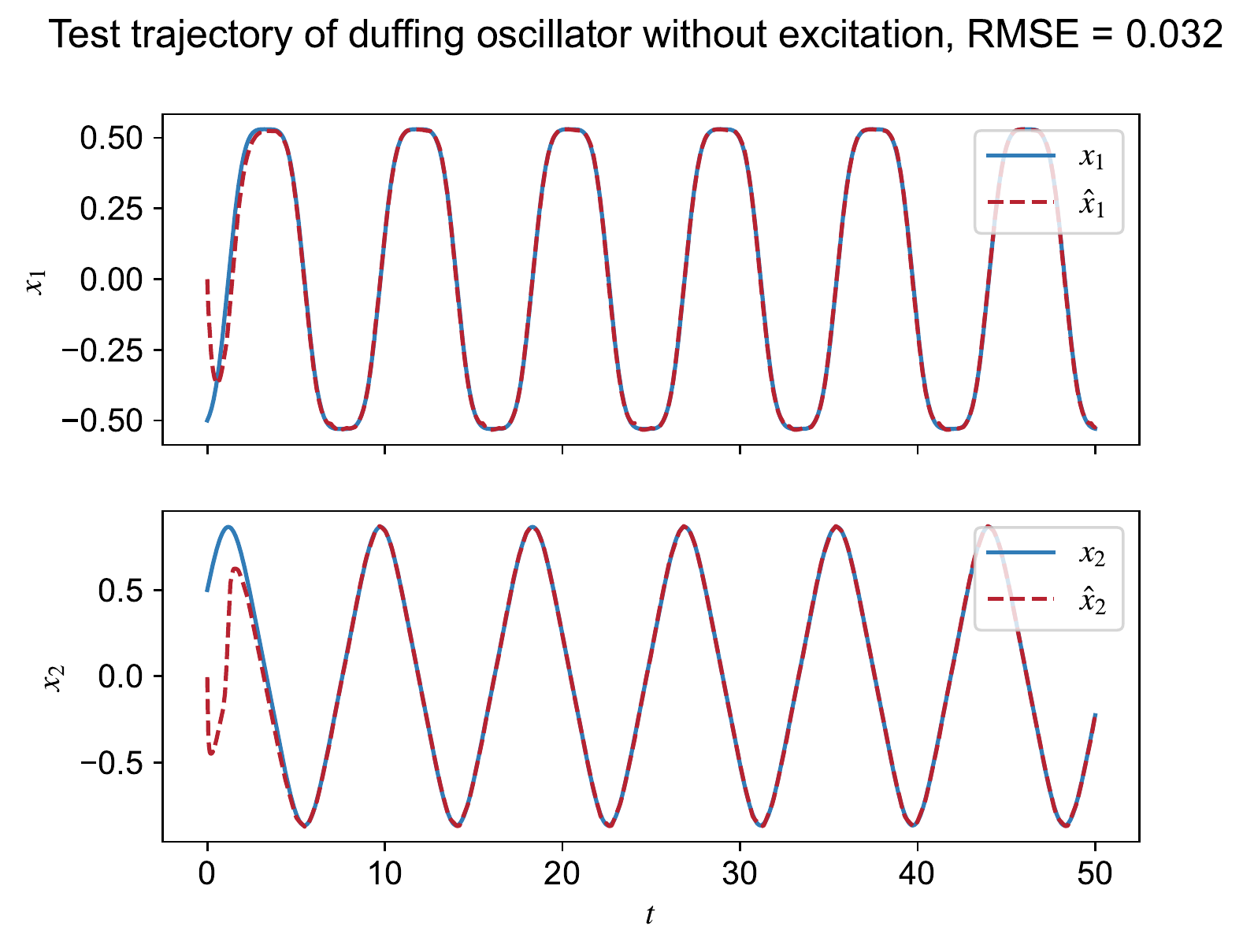}
  \caption{Test trajectory of duffing oscillator for $x_0 = (-0.5, 0.5)$ without excitation}
  \label{osci-auto}
\end{figure}

\begin{figure}[ht]
  \centering
  \includegraphics[scale=0.5]{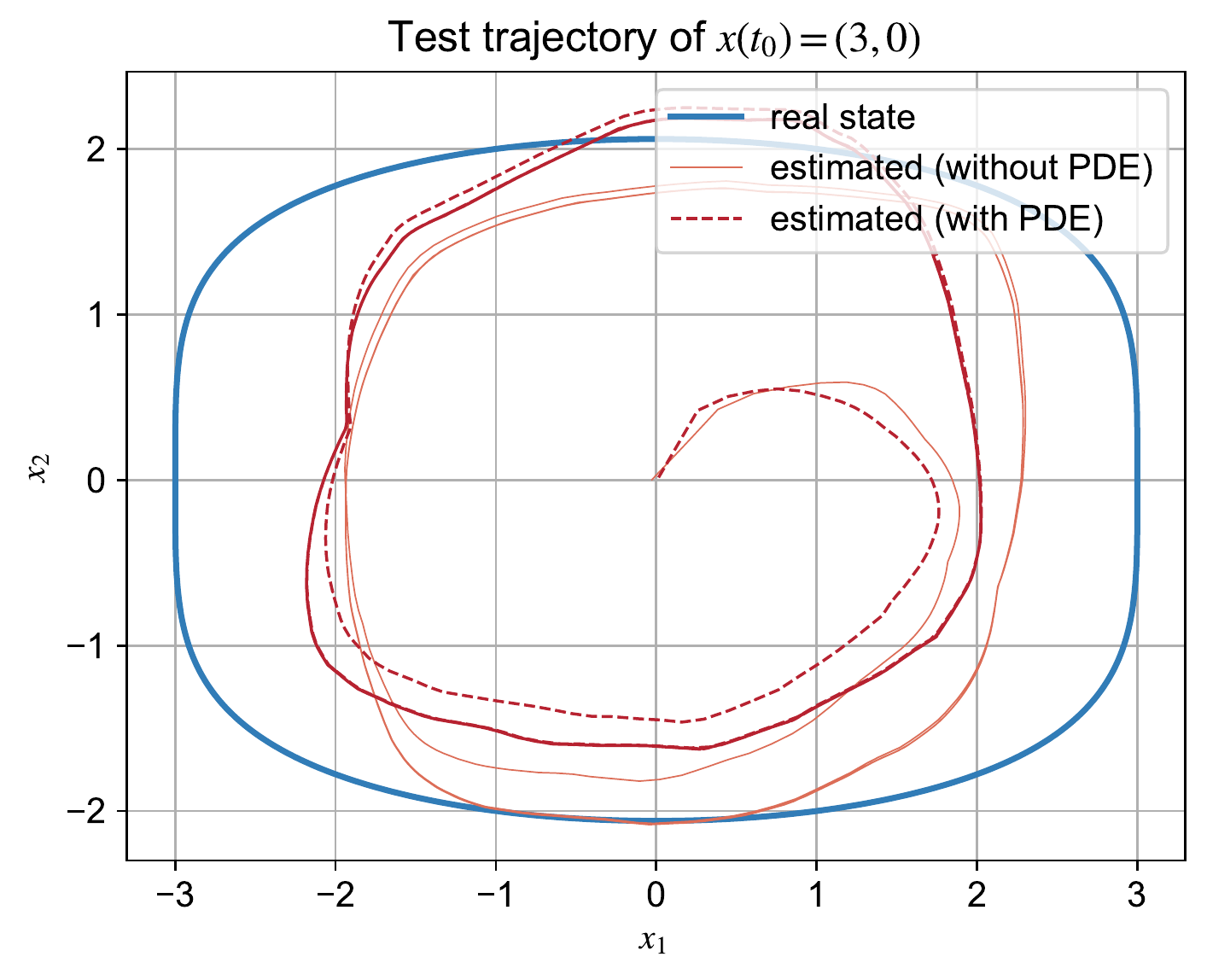}
  \caption{Test trajectory of duffing oscillator for $x_0 = (-3, 0)$}
  \label{compare_pde}
\end{figure}

\paragraph{Network Structure}
In this task, the network for $\mathcal{T}^*$ and $\mathcal{T}$ are both multilayer perceptrons with four hidden layers of 50 neurons each. $D$ is a linear layer.
\paragraph{Training Details}
The training data (initial state) is generated from a uniform random distribution $\mathcal{X}_1 = \left[-1, 1\right] \times \left[-1,1\right]$ and $\mathcal{X}_2 = \left[-3, 3\right] \times \left[-3, 3\right]$, and the dynamics are solved over the time interval $\left[0, t_f\right]$ with $t_f = 50s$. The ODE solver for Neural ODEs is chosen as RK4 with step size 0.02s. The training has been carried out for 1000 epoches using Adam optimizer with learning rate $10^{-3}$ and ExponentialLR with weight decay $10^{-4}$, batch size 50.
\paragraph{Additional Results}
The test result of learned observer for reverse duffing oscillator without excitation is shown in Figure \ref{osci-auto}.\\
\\
To incorporate the PDE constraint to improve generalization capability, the knowledge of dynamics is needed to be obtained as the additional penalty added in the loss function is $\lvert \frac{\partial \mathcal{T}}{\partial x}(x)f(x) - \left(D\mathcal{T}(x) + Fy\right)\rvert$ and this also requires that mapping $\mathcal{T}$ to be learned together with $\mathcal{T}^*$. The cases that PDE constraint is not incorporated and incorporated are trained in the same setting that $\mathcal X = \left[-1, 1\right] \times \left[-1,1\right]$. The test trajectory for the initial condition $x_0=(3,0)$ is shown in Figure \ref{compare_pde}. And RMSE of this two observers are $0.98$ and $0.86$. Since the improvement in generalization performance is not significant, and incorporating PDE constraint would also require strong knowledge of the dynamics of the system and considering the time consuming problem of computing Jacobian matrices, we did not incorporate PDE constraint in our approach.


\end{document}